\documentclass[]{aastex631}

\newcommand{\new}[1]{#1}
\newcommand{\old}[1]{}

\submitjournal{Astrophysical Journal; Accepted 23-Dec-2024}

\begin{document}

\title{A Simplified Theory of External Occulters for Solar Coronagraphs}

\author[0000-0002-7164-2786]{Craig. E. DeForest}
\affiliation{Southwest Research Institute \\
1301 Walnut St., Suite 400 \\
Boulder, CO, 80302, USA}

\author[0000-0001-6028-1703]{Nicholas F. Erickson}
\affiliation{Southwest Research Institute \\
1301 Walnut St., Suite 400 \\
Boulder, CO, 80302, USA}

\author[0009-0004-0646-0523]{Matthew N. Beasley}
\affiliation{Southwest Research Institute \\
1301 Walnut St., Suite 400 \\
Boulder, CO, 80302, USA}

\author[0009-0000-0848-5805]{Steven N. Osterman}
\affiliation{Southwest Research Institute \\
1301 Walnut St., Suite 400 \\
Boulder, CO, 80302, USA}

\author[0009-0004-5965-5498]{Travis J. Smith}
\affiliation{Southwest Research Institute \\
1301 Walnut St., Suite 400 \\
Boulder, CO, 80302, USA}

\author[0009-0005-8304-4037]{Mary H. Hanson}
\affiliation{Southwest Research Institute \\
1301 Walnut St., Suite 400 \\
Boulder, CO, 80302, USA}

\begin{abstract}

We present 
a first-principles analytic treatment of modern multi-vane occulters 
in circular (coronagraph) and linear (heliospheric imager) geometry, develop a
simplified theory that is useful for designing and predicting
their performance, explain certain visual artifacts, and explore the performance limits 
of multi-vane occulters.  
Multi-vane occulters are challenging to design in part because they violate
the conditions for both the Fraunhofer and Fresnel approximations to 
diffraction theory, and new designs have therefore generally required explicit simulation, 
empirical measurement, ``guesstimation'', or all three.  
Starting from the Kirchoff diffraction integral,
we develop
a ``sequential plane wave'' approximate analytic theory that is 
suitable for predicting performance of multi-vane
occulters, and use it to derive closed-form expressions for the performance
of new designs.
We review the fundamental 2-D system of an occulter edge, discuss how it 
applies to real 3-D systems by extrusion or revolution, 
present the reason for observed bright quasi-achromatic fringing around 
coronagraph occulters, develop the sequential plane wave
approximation in 2-D and explore its limits,
describe the relevance of the 2-D theory to practical 3-D instruments, 
and discuss implications for multi-vane
occulter design in current and future instruments.
\end{abstract}


\section{Introduction} \label{sec:intro}

The solar corona, viewed in visible light, has roughly one millionth the total intensity of the solar photosphere.  The contrast is so large that
the corona is not normally visible without a total solar eclipse.  Imaging
the corona under normal conditions requires specialized
equipment to reject light from the much brighter solar photosphere. The
technology of specific coronagraph instruments
has been described by many authors as the technology has developed over the past
century 
\citep[e.g.,][]{lyot_1930,Evans1948,newkirk_bohlin_1963,koomen_etal_1975,brueckner_etal_1995,howard_etal_2008,yang_etal_2018,deforest_etal_2022}. 
The early history and basic theory of coronal imaging, polarization
measurement, and stray light rejection are covered by \citet{Billings1966}. 
More recently, ``heliospheric imagers'', which work on the same principles as
coronagraphs but observe farther from the Sun, have proved useful and require
even deeper stray light rejection
\citep[e.g.,][]{Jackson_1985,howard_etal_2008,vourlidas2016,howard2020,deforest_etal_2022}. 
The current generation of heliospheric imager uses imaging and stray-light 
rejection strategies similar those of coronagraphs, but typically in linear geometry
rather than circular geometry.

Instruments to image the lower solar corona typically reject the solar light in an
image plane: a low-scatter optic is used to develop an image of the Sun itself 
and the surrounding corona, and the solar light is either reflected/obstructed
by an object at the focal plane \citep[e.g.,][]{dewijn_2012}, or allowed to 
exit the instrument through a hole in a mirror 
\citep[e.g.,][]{brueckner_etal_1995}.  \new{This provides a sharp boundary on the image plane, separating photospheric light from coronal light without a wide vignetted boundary zone; but 
it also requires very low scatter in the first optic, which must minimize scattering of the bright photospheric beam into the much fainter coronal beam.}

Because the coronal brightness drops off very rapidly, rejecting photospheric
light from a focused image is not feasible for heliospheric imagers or for
wide-field coronagraphs that view 
more than a fraction of a degree from the limb of the Sun itself.  
Solar coronagraphs viewing several solar radii from the Sun must reject
stray light from the photosphere, by a factor smaller than $10^{-11}$; and
wider angles require greater rejection.  In this regime, external occultation
is critical: the first optical element of the instrument is an occulting body that 
rejects the photospheric light via reflection or absorption, to separate it 
from the incoming beam before it can be scattered within the imaging system.  The 
physics of diffraction around the occulter determine the performance
of the instrument, and dominate the design of space instrumentation.

The basic theory of diffraction follows straightforwardly from the wave theory of 
light and is covered in 
common textbooks \citep[e.g.,][]{feynman1963,hecht_zajac_1974,bornwolf1999}; 
but a general treatment
is intractable for analytic methods for all but the simplest cases.  The two most famous
simplified cases are Fraunhofer diffraction, in which the Born approximation holds and 
phase delays are considered to be linear in both incident angle and offset within the 
instrument; and Fresnel diffraction, in which phase delay is allowed to be quadratic in either 
incident angle or position. Although Fresnel diffraction is very useful in special cases, it
is limited in its generality for occulter design, because many occulter
solutions are three-dimensional objects and do not adapt well to the 
quadratic-phase-on-a-plane approach of the Fresnel approximation.  \new{Proper
 treatment of multiple vanes generally requires evaluating the Kirchoff integral
 \citep[e.g.,][Chap. 8]{bornwolf1999} with little or no approximation, which in turn 
 requires numerical evaluation/modeling; software tools used
for this purpose include FRED \citep{photonengr} and VirtualLab \citep{lighttrans}. 
In practice, up to two-vane occulters are tractable for analytic treatment, but 
systems with three and more vanes are not\citep{aime2020}.}

The problem of stray light rejection in a solar instrument is specific enough
that general methods often do not apply \new{straightforwardly}.  As a result, treatment of solar
occulter systems is quite varied.  Several studies have focused on 
performance of jagged or anti-aliased single occulters 
\citep{newkirk_bohlin_1963,fort_etal_1978,lenskii_1981,shirley_datla_1996,aime_2007}; 
these may be treated with the Fresnel approximation and we do not
consider them here.  Round multi-disk external occulters have proved more tractable 
to fabricate and align, and have been incorporated in many prototype and 
actual instruments, from initial studies by \citet{newkirk_bohlin_1963}, 
through the LASCO instruments \citep{brueckner_etal_1995,bout_etal_2000}, to
the more modern STEREO/COR \citep{gong_socker_2004}, 
SOLO/METIS \citep{landini_etal_2012}, ISS/CODEX \citep{yang_etal_2018}, 
 GOES/CCOR \citep{Thernisien_2021}, and the ground-based CATEcor \citep{deforest2024catecor}.  The design analysis methods for these
instruments have been as varied as the occulters themselves, with a 
significant focus on highly simplified \citep[e.g.,][]{newkirk_bohlin_1963},
intuitive \citep[e.g.,][]{buffington2000}, empirical 
\citep[e.g.,][]{bout_etal_2000}, and proprietary numerical 
\citep[e.g.,][]{gong_socker_2004} methods.  More recent work has
explored the 
practical limits of analytic treatment \citep[e.g.,][]{aime2020,wang_etal_2021}.

In considering design principles for future coronagraphs and 
heliospheric imagers, we sought to develop
a middle ground: broadly applicable analytic design rules simpler to
evaluate than the full
Kirchoff integrals, which nevertheless provide both more rigor than 
intuitive empirical ``horse sense'', and also well-defined limits of 
applicability. 
In this article, we develop from first principles a sufficient, simplified
theory to understand and design externally occulted coronagraphs and
heliospheric 
imagers, including why and how multi-edge occulters work in circular and 
linear geometry.  We develop a 2-D approximation (successive-plane-wave
approximation, SPW) for designing multi-edge
occulters, derive from it the ideal geometry 
(cylindrical or ogive envelope) for a 
multi-edge system, and explore
the limits of the SPW approach.  As corollaries we derive 
both the root cause for
the achromatic fringing
seen around the occulters of existing externally occulted 
coronagraphs, and also 
the limits on
performance of multi-edge occulters and their $n\rightarrow\infty$ limit, smooth occulters.  The 2-D theory applies directly to extruded designs in 3-D,
such as are used for contemporary heliospheric imagers \citep{howard_etal_2008, vourlidas2016, laurent_etal_2025}; we also show how 
it applies to design of revolved designs in 3-D, such as in externally occulted coronagraphs.

Section \ref{sec:half-plane} is an abbreviated derivation of the basic
Fresnel attenuation formula to describe illumination inside the 
umbra of a sharp shadow mask (occulter). 
Section \ref{sec:fringes} develops the 
angular distribution of Fresnel-diffracted light around the edge, and shows the
origin of the ``bright ring'' and ``bright line'' fringes that are seen in existing 
coronagraph and heliospheric 
imager data \citep[e.g.,][]{howard_etal_2008}. 
Section \ref{sec:multiple} expands the theory 
to modern multi-edge occulters via an expedient approximation (the successive-plane-wave
\new{approximation}), develops a design theory for curved-envelope multi-edge occulters, 
briefly touches on smooth occulters as a limiting case of multi-edge occulters, and
discusses breakdown of the successive-plane-wave approximation.
Section \ref{sec:disks} develops a theory
of multi-disk coronagraph occulters as a perturbation of the 2-D theory (which is
also directly applicable to linear occulters as in heliospheric imagers).  
Finally, 
in Section \ref{sec:discussion-conclusions} we summarize the key points that 
develop from the theoretical discussion, and relate them to existing
and future coronagraphs.

\section{A quick review: 2-D Fresnel diffraction around an edge\label{sec:half-plane}}

\begin{figure}
    \centering
    \includegraphics{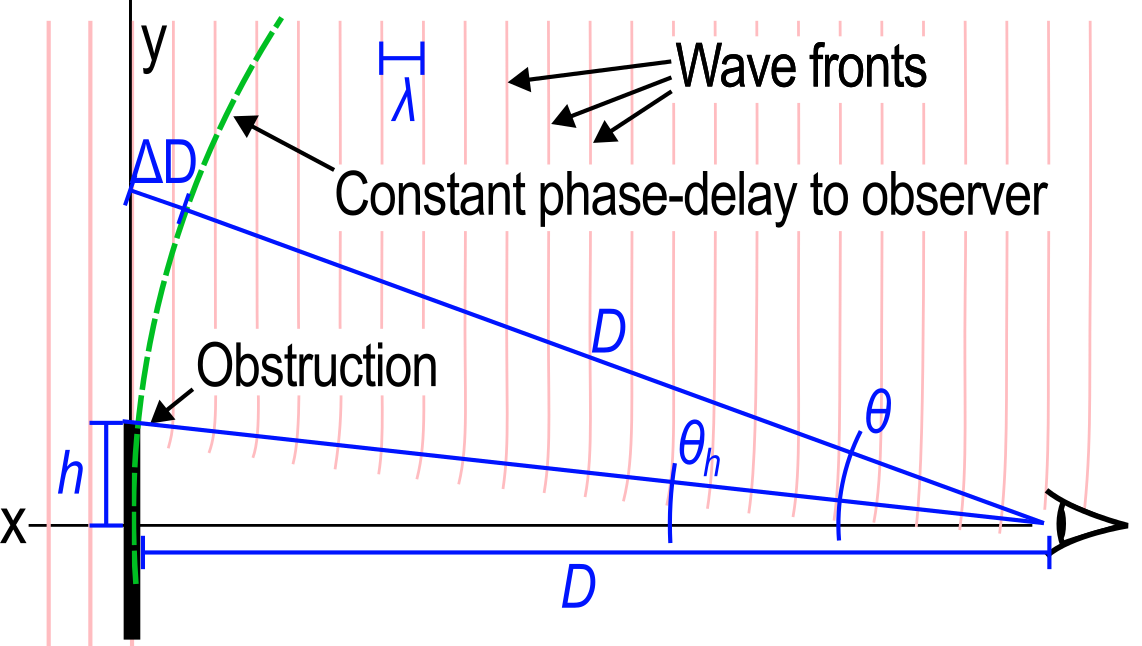}
    \caption{Fresnel scattering limits occulter performance.  The most basic Fresnel calculation in 2-D involves plane waves incident on a simple occulting body.  Constant-phase-delay surfaces around an observer at distance $d$ from the body form circles on the plane.  Each line of sight from the observer imposes a phase delay $\Delta D/\lambda$  In the Fresnel approximation, $\Delta D \propto \theta^2$.  }
    \label{fig:fresnel-geometry}
\end{figure}

The Fresnel approximation to diffraction covers systems in which phase delay $\Delta \phi$
for the 
constituent paths of a diffracted wave is quadratically dependent on 
location in the scattering body; this contrasts with the even more common 
Fraunhofer/Born 
approximation, in which $\Delta \phi$ is linearly dependent on location.
While treatments of Fresnel diffraction may be quite complex (and form the basis for
photon sieves, Fresnel focusing optics, and holography), the simplest
case (applicable to solar occulters) may be treated
in 2-D (Figure \ref{fig:fresnel-geometry}). This essential 2-D treatment is
directly applicable to heliospheric imagers, in which the
system is extruded into the third dimension \citep[e.g.,][]{howard_etal_2008}, and 
may be adapted to form a theory of externally occulted coronagraphs, in which the system
is revolved around an axis of symmetry \citep[e.g.,][]{brueckner_etal_1995}.  While 
the problem of Fresnel diffraction around a simple occulting body is covered in some 
detail in standard optics textbooks \citep[e.g.,][, Chapters 8-11]{bornwolf1999}, it is worth 
an abbreviated treatment to establish terms and build immediate intuition.

Figure \ref{fig:fresnel-geometry}
sets up the essential geometry of the Fresnel diffraction problem around a
simple occulter, which may be considered as an infinite half-plane extending into 
(and out of) the
page in 3-D.  
The obstruction, of height $h$, is a distance $D$ from the
observer, and the locus of constant phase-delay from the incoming waves to the
observer is a circle of radius $D$. 
Incident waves passing through the $y$ axis below
$y=h$ are truncated by the obstruction.  Those above $y=h$ are subjected to a phase
delay from the difference between the constant-phase-delay surface and the $y$ axis.
The additional distance $\Delta D$ results in a phase delay of $\Delta D / \lambda$ 
cycles (i.e. $2\pi \Delta D / \lambda$ radians).  The wave fronts are taken to travel
along the $x$ direction, but without loss of generality as the angle of the obstruction
relative to the wave fronts is unimportant to the problem.

The analysis is set up following Huygens' construction, replacing
the incident plane wave with a line of oscillating sources along the $y$ axis. 
Circular expanding wave fronts from each point on the axis interfere at the observer,
with a phase delay that depends on the $y$ coordinate and is determined by the distance
$\Delta D$ normalized by the wavelength $\lambda$.

When considering Fraunhofer diffraction, for example in a single-slit or two-slit 
problem, one treats the constant phase-delay surface as piecewise flat, so that 
$\Delta D \propto y - y_0$ for some $y_0$ that fits the problem geometry.  In the
full wave treatment, 
\begin{equation}
    \new{\Delta D = \sqrt{ D^2 + y^2 } - D}\,,\label{eq:full-geometry}
\end{equation}
which is analytically intractable.  The Fresnel approximation 
expands the square root and keeps terms only up to $y^2/D^2$ (i.e. $\theta^2$) 
approximation order, i.e.
\begin{equation}
    \Delta D = D \left({y^2}/{2D^2} + ... \right)\,\approx y^2/2D = D\theta^2/2.\label{eq:Fresnel-geometry}
\end{equation}

With the approximation that $y \ll D$, projection polarization effects are 
negligible and light
may be treated as a scalar wave (neglecting polarization and treating the 
electric field as a scalar quantity). Similarly, the radius 
$r\equiv\sqrt{y^2+D^2}$ is independent 
of $y$ in this regime and may be replaced with $D$, so that the 
$1/r$ dependence of the oscillating electric field on distance from 
a source may be neglected.  
Invoking Huygens' principle and replacing the oscillating 
electric field of an incident wave with a matching collection of oscillating sources, we can follow Fresnel and 
immediately write:
\begin{equation}
    E_{obs} = \frac{\phi E_0}{\sqrt{D\lambda}} \int_{h}^{\infty} 
    e^{i y^2 \pi/D\lambda}
    dy\,\label{eq:fresnel-integral}
\end{equation}
where the (complex-valued)
$E_0$ is the oscillating field at the $y$ axis from the incoming
plane wave, such that 
$E_0 E_0^* = I_0$, the incident
intensity; $\phi$ is a constant complex phase-delay factor (i.e. $\left|\phi\right|$ = 1) that is unimportant to the overall intensity; 
$D$ and $h$ are as
shown 
in Figure \ref{fig:fresnel-geometry}; the $\sqrt{D\lambda}$ denominator normalizes for distance 
in the 
2-D geometry and the required Huygens' wavelet normalization by $\lambda$ \citep[e.g.,][ch. 8]{bornwolf1999}; 
the integral is over all points on the $y$ axis 
above the obstruction; the imaginary exponential tracks the phase delay introduced 
by $\Delta D$ at 
each value of $y$; and $\lambda$ is the wavelength of the incoming plane wave.
Equation \ref{eq:fresnel-integral} is the Fresnel approximation to the oscillating electric field at 
the observer in Figure \ref{fig:fresnel-geometry}, due to equivalent source motion
of a family of oscillating electric sources along the $y$ axis to ``mock up'' the 
incident plane wave shown at left of Figure \ref{fig:fresnel-geometry}.

The integral in Equation \ref{eq:fresnel-integral} does not include a 1/$r$ term, 
because $r\approx D$ when $y$ is small, and the constant $1/D$ has been brought 
outside the integral.  Integration to $+\infty$ breaks the $y\ll D$ approximation,
but may be
justified in two ways: analytically, by noting that in places where the
approximation doesn't hold, the
integrand oscillates very quickly, and therefore does not contribute significantly to the
overall integral;
and observationally, by noting that in real optical systems, occulter stray
light is imaged mainly into a bright locus close to the occulter, i.e. where 
$y\ll D$ 
(Section \ref{sec:fringes}). A more rigorous treatment of this approximation may be found in
\citet{bornwolf1999}, Chapter 11.  \new{Making the jump from 2-D to 3-D involves 
either extrusion along the $z$ (out-of-plane) axis for a linear baffle/occulter,
or revolution about the $x$ axis for a circular occulter.  The first case is 
straightforward as $z$ doesn't participate in Equation \ref{eq:fresnel-integral}.
The second requires some care and is discussed in Section \ref{sec:disks}.}

\begin{figure}
    \centering
    \includegraphics {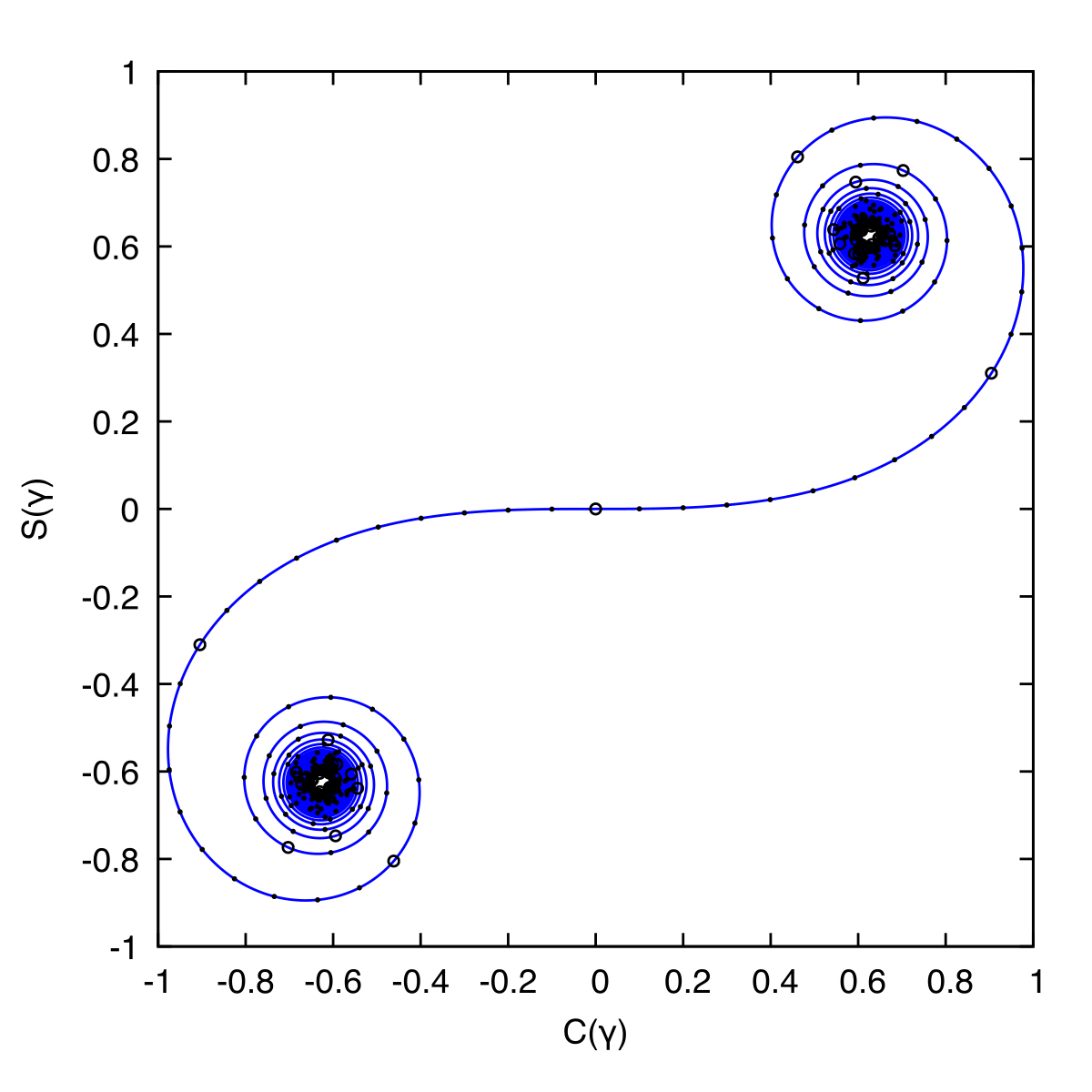}
    \caption{The Cornu spiral graphically represents the real and imaginary parts $\mathcal{C}(\gamma)$ and $\mathcal{S}(\gamma)$, respectively, of the Fresnel integral.  The $\gamma$ parameter is 0 at the origin.  Integer values of $\gamma$ are marked with small circles; every 0.1 is marked with a tick.  The two asymptotes are at $\pm \sqrt{\pi/8}$ for both $\mathcal{C}(\gamma)$ and $\mathcal{S}(\gamma)$.  A line drawn between two points on the spiral represents the complex Fresnel integral between the two corresponding values of $\gamma$, and its squared length is proportional to the corresponding intensity of light.}
    \label{fig:cornu-spiral}
\end{figure}

The integral in Equation \ref{eq:fresnel-integral} can be simplified considerably.  Picking the
characteristic length
\begin{equation}
    s \equiv \sqrt{D\lambda/\pi}\,,
    \label{eq:define-s}
\end{equation}
and then defining a new variable $u$, such that $y=su$, yields
\begin{equation}
    E_{obs} = \frac{\phi E_0}{\sqrt{\pi}}\int_{h/s}^{\infty}e^{iu^2}du\,.
    \label{eq:simpler}
\end{equation}
The integral is transcendental and is described with the special functions 
\begin{equation}
    \mathcal{C}(\gamma) \equiv \int_0^\gamma \cos(u^2) du\,.\label{eq:C-function}
\end{equation}
for the real part and 
\begin{equation}
    \mathcal{S}(\gamma) \equiv \int_0^\gamma \sin(u^2) du\,\label{eq:S-function}
\end{equation}
for the imaginary part.  Both functions are odd, and both approach the value $\pm\sqrt{\pi/8}$ as
$\gamma \to \pm\infty$.  Plotting them
parametrically vs. $\gamma$ on the complex plane yields the famous ``Cornu spiral'' in 
Figure \ref{fig:cornu-spiral}, which graphically represents definite integrals of the form in 
Equation \ref{eq:simpler}, without the scaling coefficients in front, and therefore forms a useful nomogram for evaluating integrals with the form of Equation \ref{eq:simpler}.  Important properties of the Cornu spiral include that its path length from the 
origin is equal to the $\gamma$ parameter, that it slowly approaches limit points 
as $\gamma \rightarrow \pm\infty$, and that the local curvature of the spiral is proportional 
to $\gamma$.  The curvature property was used in architecture and technical drawing throughout 
the pre-digital drafting era, in the form of the ``French curve'' templates that may be used 
to render smooth curves of arbitrary radius.

Writing the integral in terms of $\mathcal{C}$ and $\mathcal{S}$ puts a name to its solution,
although the actual calculation of values on the Cornu spiral is typically via successive approximation.  $E_{obs}$ is given by
\begin{equation}
    E_{obs} = \frac{\phi E_0}{\sqrt{\pi}} \left\{
    \left[ \sqrt{\pi/8} - \mathcal{C}(h/s) \right] +
    \left[ \sqrt{\pi/8} - \mathcal{S}(h/s) \right] i
    \right\}\,,\label{eq:e-field}
\end{equation}
and (remembering that $\phi^*\phi$=1) the intensity  $I_{obs}$ at the observer is therefore
\begin{equation}
    I_{obs} = I_0 \frac{1}{\pi} \left\{
    \left[ \sqrt{\pi/8} - \mathcal{C}(h/s) \right]^2 +
    \left[ \sqrt{\pi/8} - \mathcal{S}(h/s) \right]^2
    \right\}\,,
    \label{eq:intensity}
\end{equation}
or, in terms of the angle $\theta_h$ and eliminating $s$,
\begin{equation}
    I_{obs} = I_0 \frac{1}{\pi} \left\{
    \left[ \sqrt{\pi/8} - \mathcal{C}\left(\theta_h \sqrt{D/\pi\lambda}\right) \right]^2 +
    \left[ \sqrt{\pi/8} - \mathcal{S}\left(\theta_h \sqrt{D/\pi\lambda}\right) \right]^2
    \right\}\,,
    \label{eq:intensity-by-angle}
\end{equation}
where $\theta_h$ is the bend angle of the light at the obstruction (as in Figure \ref{fig:fresnel-geometry}) and $s$ 
is the characteristic length given by Equation \ref{eq:define-s}.

\begin{figure}
    \centering
    \includegraphics{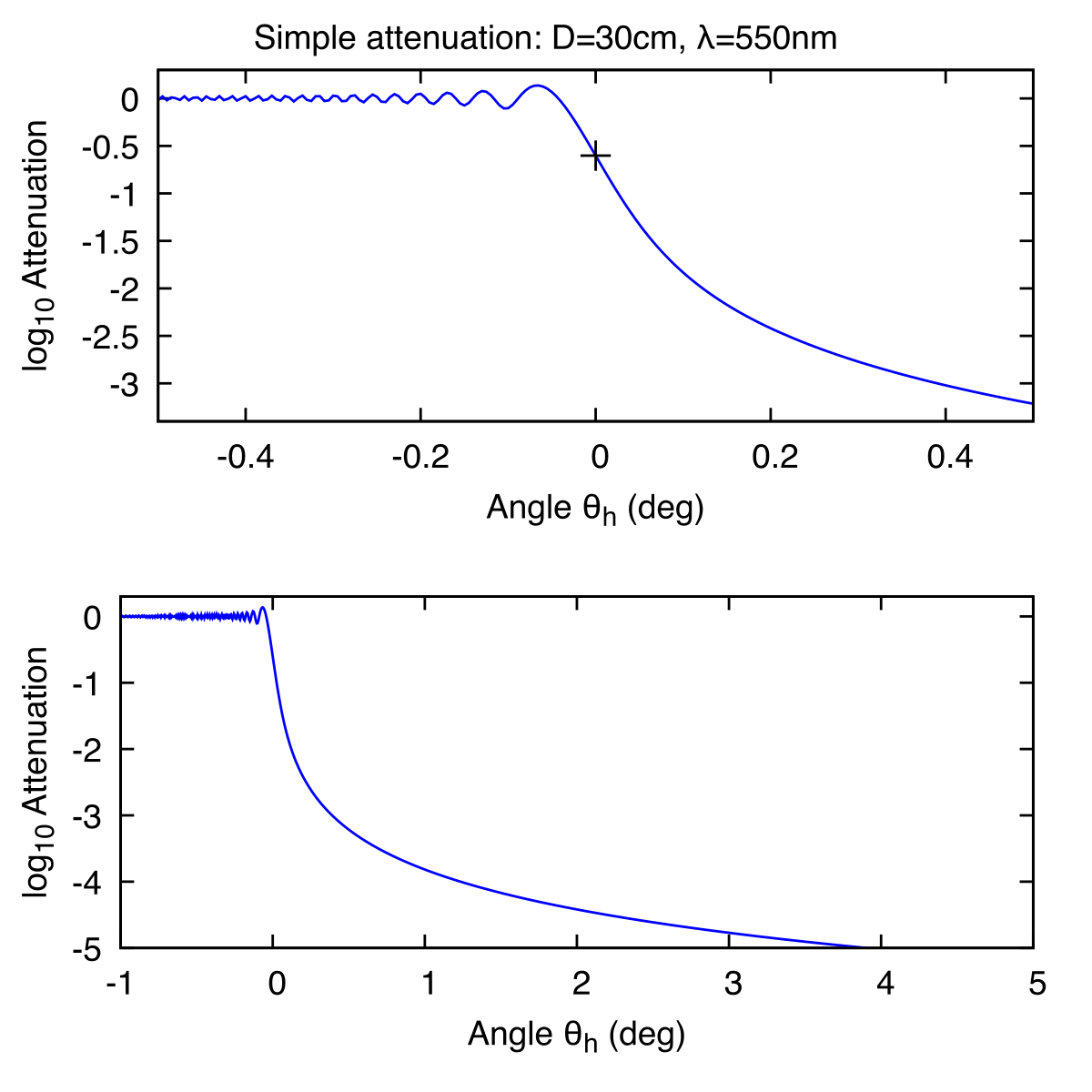}
    \caption{Attenuation from a sample single-edge occulter, calculated from Equation \ref{eq:intensity-by-angle} for $\lambda$=550nm and $D$=250mm, reveals the major aspects of Fresnel scatter by a half-plane: attenuation coefficient at $\theta_h$=0 is 
    0.25 (i.e. -0.6 on the log$_{10}$ scale) and is marked with ``+''; high negative bend angles admit all light; low negative bend angles cause interference; and positive bend angles
    attenuate the light rapidly.  Top plot shows attenuation close to the conventional edge of the occulter's shadow.  Bottom plot shows wider bend angles.}
    \label{fig:intensity-curve}
\end{figure}

Performing the Fresnel integrals is a matter of general interest and is relevant to many fields including
radio-link engineering, holography, spectral and coded imaging, phase-shift imaging, and even cryptography.  More
prosaically, Equation \ref{eq:intensity-by-angle} is useful for coronagraph design.  It gives the total beam 
attenuation at a point in the umbra of a simple
obstruction occulter from a monochromatic point source at infinity (collimated beam), given the wavelength
of the light, distance from the occulter to the observer, and required bend angle of the beam to get around 
the edge of the occulter.  
The attenuation is plotted in Figure \ref{fig:intensity-curve} for some example values of $\lambda$, $D$, and $\theta_h$, 
and reproduces textbook half-plane attenuation formulae \citep{hecht_zajac_1974,bornwolf1999}.

Although Equation \ref{eq:intensity-by-angle} treats only a monochromatic beam and point source, one can integrate it
numerically over angle and wavelength, to compute scattering attenuation from distributed and polychromatic sources. \new{We discuss distributed sources, for the circular case only, in Section \ref{sec:disks}.}

Exploring the limits of Equation \ref{eq:intensity} yields intuition.  When $h\to-\infty$,
the definite integral extends between the two asymptotes in Figure \ref{fig:cornu-spiral}, 
and the expression collapses to $I_{obs}=I_0$ as it should with no obstruction present.
When $h\to+\infty$, the definite integral extends from the positive infinity asymptote to itself, and 
the expression collapses to $I_{obs} = 0$ as it should when the beam is fully obstructed.  
When $h=0$, $I_{obs}=I_0/4$, a famous result from Fresnel that helped (along with the presence of the Arago spot) to cement adoption of the wave theory of light \citep{arago1819}.

The form of the arguments to $\mathcal{C}$ and $\mathcal{S}$ in 
Equation \ref{eq:intensity-by-angle} highlights a design
constraint on physical geometry of occulters.
For a given attenuation coefficient $I_{obs}/I_0$, 
\begin{equation}
    \theta_h \sqrt{D/\lambda}=k\label{eq:design-constraint}
\end{equation}
where $\theta_h$ is the angle between the optics and occulter edge in
a coronagraph, also called the ``inner field of view cutoff'' \citep{Thernisien_2005}; and $k$ is a constant that depends on the particular value 
of $I_{obs}/I_0$.  This is the essential design trade
for externally occulted instruments: longer instruments can image closer
to the Sun (by minimizing $\theta_h$) -- but doubling the length of the
instrument only achieves a roughly 30\% decrease in the required angular margin between
the limb of the Sun and the inner field of view cutoff.

\section{Imaging Fresnel-diffracted light\label{sec:fringes}}

\begin{figure}
    \centering
    \includegraphics{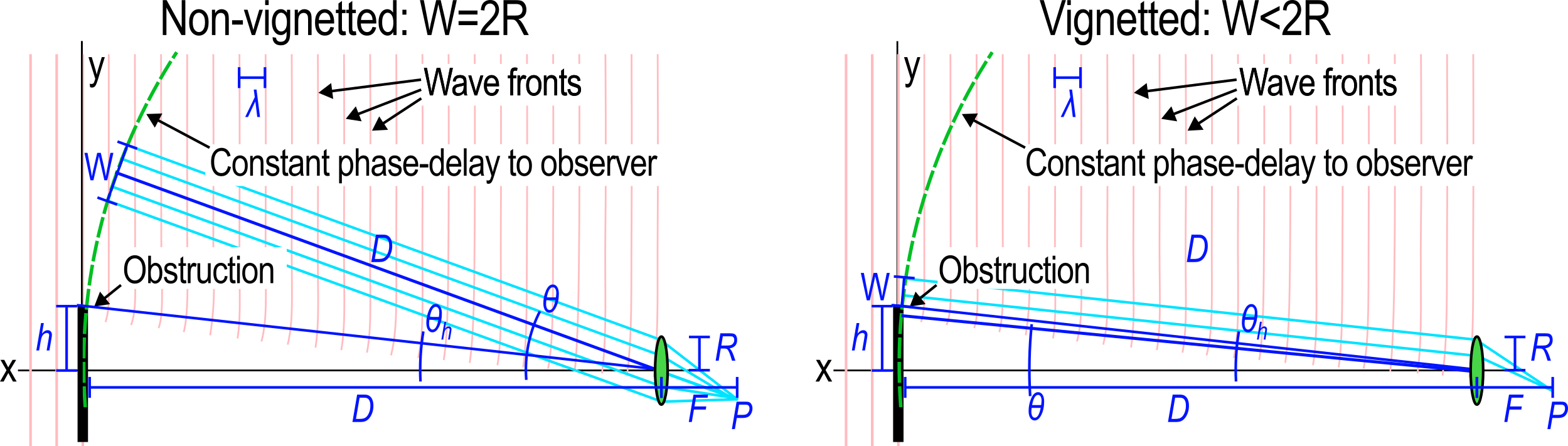}
    \caption{Geometry of an imaging system is different than that of a simple 
    intensity sensor. The 
    aperture of a focusing optical system forms a ``virtual slit'' of width $W$
    on the 
    constant-phase surface of Figure \ref{fig:fresnel-geometry}, imposing a 
    Fraunhofer diffraction
    pattern on the image plane.  At wide angles (apparent distances) from the occulter (left panel), W is set by the
    entrance aperture and the Fraunhofer 
    diffraction pattern takes the standard form in Equation 
    \ref{eq:evaluate-sinc}. At narrow angles (apparent distances) close to the occulter (right panel), W depends on the
    relative geometry of the beam and the occulter, modifying the functional 
    form as in Equation \ref{eq:vignetted-intensity}.}
    \label{fig:fringe-geometry}
\end{figure}

Coronagraphs and heliospheric imagers are imaging systems that 
resolve angle, so although an integral over the $y$ axis (or equivalently $\theta$) can predict the total amount of diffracted
light
entering the aperture, it cannot identify the diffraction features seen by a coronagraph or heliospheric imager. 
For 
that, it is
necessary to fix $\theta$ (for a given pixel) and carry out the integral across the
aperture instead of across incident angle.  Figure \ref{fig:fringe-geometry} illustrates a
simple
imaging system with focal
length $F$ and aperture radius $R$.  Light entering the aperture at angle $\theta$ is focused onto a focal
point $P$ behind the imaging system, a distance $\theta F$ below the $x$ axis. 

\begin{figure}
    \centering
    \includegraphics[width=6in]{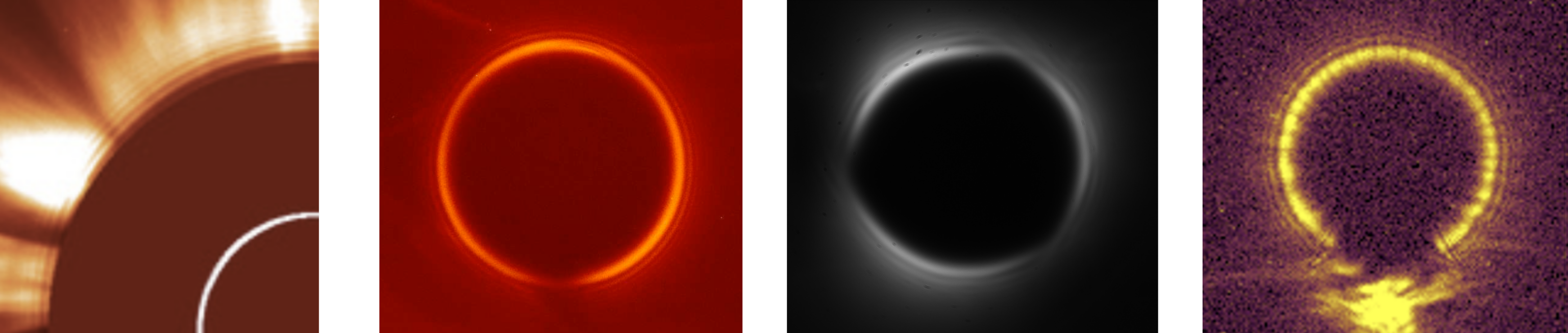}
    \caption{Quasi-achromatic fringing is characteristic of external occultation as seen with four separate externally occulted instruments. Each image shows fringing around the image of the external occulter in the focal plane, following Equation \ref{eq:fraunhofer-spacing-little-theta0}.  (A) SOHO/LASCO C2 \citep{brueckner_etal_1995}, though the lowest order fringes are obscured by C2's internal occulter; (B) A conical occulter under test \citep{yang_etal_2018}; (C) CATEcor \citep{deforest2024catecor}, (D) SwSCOR \citep{Erickson_etal_2025}.  The LASCO image is zoomed in to reveal the high-order fringes from the external occulter, as seen around the internal occulter.}
    \label{fig:fringing}
\end{figure}

At apparent locations (i.e. angles) far from the occulter,
the imaging system works essentially independently of the occulter itself.  
But at angles close to the inner portion of the field of view, the aperture 
is vignetted by the occulter and the effective aperture of the instrument is 
reduced.  This reduces total effective area of the instrument, and also 
affects the point-spread function due to conventional Fraunhofer diffraction
through the reduced effective aperture \citet[e.g.,][]{Llebaria_etal_2006}.
But the reduction of effective aperture size also affects the stray light
pattern itself.  Although straightforward models of coronagraph performance 
tend to reveal a ``bright ring'' around the occulter \citep[e.g.,][]{Rougeot_etal_2018}, real externally occulted coronagraphs exhibit 
diffraction fringes around the occulter, even when the optics are operated with
broadband light.  Some examples are shown in Figure \ref{fig:fringing}.

To understand these imaged stray light features, one must consider 
more deeply the focusing system after the occulter.
In Figure \ref{fig:fringe-geometry}, the focusing system holds constant the 
phase delay between a given pixel on the detector and all the points on a 
flat (or, in the 2-D treatment, linear) ``virtual slit'' of width $W=2R$ 
(with the same smallness
condition as in 
Section \ref{sec:half-plane}, i.e. $y\ll D$ or equivalently $\theta\ll 1$).
Because the virtual slit is canted (by $\theta$) from the incident wavefronts and is small compared to the curvature of the constant
phase surface to the aperture, the resulting
intensity at $P$ is subject to conventional Fraunhofer diffraction: there is an approximately linear relationship between incoming phase and position across the virtual slit.  Revisiting the setup of Equation \ref{eq:fresnel-integral}, with (for convenience) a square aperture $2R$ on a side,
\begin{equation}
    E_{P} = \frac{\phi E_0  R}{DF\sqrt{\lambda}}  \int_{Q}^{R} e^{i\frac{2\pi}{\lambda}\theta y'}dy'\,,
\end{equation}
where $E_P$ is the computed electric field at Point P in Figure \ref{fig:fringing}, $\phi$ is again an overall phase factor; $E_0$ is the incident wave electric field (treated 
as a scalar, as in Section \ref{sec:half-plane}); 
the $D$ 
denominator arises from using the full 3-D geometry with linear extrusion of Figure \ref{fig:fringe-geometry} in the $z$ direction (out of the page); a factor
of $R$ in the numerator accounts for the aperture integral in the $z$ direction, over which the integrand is constant; $F$ is the 
focal length of the optics; $dy'$ is
a perturbative variable of 
integration around $y$ (i.e. centered on $D\theta$); and the lower limit of integration, $Q$,
is the greater of $-R$ and $(D\theta-R)-h$.  The integral is just the Fourier transform of
the rectangular
function from Q to R,
\begin{equation}
    E_{P} =  \frac{\phi E_0 R \sqrt{\lambda}}{i 2\pi \theta D F }
    \left[e^{i\frac{2\pi}{\lambda}\theta y'}\right]_{y'=Q}^R\,.\label{eq:sinc-virtual-slit}
\end{equation}
In the far (unvignetted) field, $Q=-R$ and Equation \ref{eq:sinc-virtual-slit} devolves
to the familiar single-slit diffraction formula for the E-field at a non-vignetted focal plane point $P_{nv}$:
\begin{equation}
    E_{Pnv} = \frac{\phi E_0 R \sqrt{\lambda}}{\pi\theta D F}\sin\left(\frac{2\pi}{\lambda}\theta R\right)\,, 
    \label{eq:non-vignetted}
\end{equation}
with the corresponding intensity
\begin{equation}
    I_{Pnv} = I_0 \frac{R^2 \lambda}{\pi^2 \theta^2 D^2 F^2} \sin^2\left( \frac{2\pi}{\lambda} \theta R \right)\,.
    \label{eq:evaluate-sinc}
\end{equation}
Equation \ref{eq:evaluate-sinc} is just the usual wide-slit Fraunhofer 
formula -- a squared sinc function in $\theta$ -- that is scaled to the specific geometry of the $2R\times2R$ aperture and focal length of the
optics, giving the intensity $I_{Pnv}$ at each location in the focal plane. 
$I_{Pnv}$ has zeroes at
\begin{equation}
    \theta = \frac{n\lambda}{2R}
\end{equation}
for every $n\in\mathbb{Z}^+$, so that each peak is half the width of the
conventional diffraction limit around the $n=0$ peak.  
Given that $n$ is large across the image, and that most instruments are at
least somewhat polychromatic, the oscillating factor disappears in a 
typical instrument, and we are left with a locally smoothed stray light
intensity $\bar{I}_{Pnv}$:
\begin{equation}
    \bar{I}_{Pnv} = I_0 \frac{R^2 \lambda}{2 \pi^2 \theta^2 D^2 F^2} ,
    \label{eq:I_Pnv-avg}
\end{equation}
which describes a smooth stray light pattern extending across the field of view and varying as $\theta^{-2}$ and $\lambda$.

Where $D\theta < h+R$, the beam intersects the occulter vane and $Q\ne -R$.
In that portion of the field of view, the cosine terms implicit in
Equation \ref{eq:sinc-virtual-slit} don't cancel directly as in
Equation \ref{eq:non-vignetted}.  But, recognizing the system's symmetry
under
displacement, we can
adopt new symmetric limits by re-centering the coordinate system on the
slit, and the cost of adopting fixed phase delay.  Setting $\theta_0$ to 
be the lowest 
angle at which any ray can enter
any part of the square aperture, 
\begin{equation}
    \theta_0 \equiv \frac{h-R}{D}\,,
    \label{eq:theta-zero}
\end{equation}
and one can re-center using an equivalent-radius $R'$.  Defining :
\begin{equation}
    R' \equiv (\theta - \theta_0) D / 2\,.
    \label{eq:R-prime}
\end{equation}
Then, at vignetted points $Pv$,
\begin{equation}
    E_{Pv} = \frac{\phi \psi E_0 R \sqrt{\lambda}}{\pi\theta D F}\sin\left(\frac{2\pi}{\lambda}\theta R'\right)\,,
    \label{eq:vignetted-sinc}
\end{equation}
where, compared to Equation \ref{eq:evaluate-sinc}, the $\psi$ is another phase-delay factor, the 
numerator factor of $R$ remains the same because it is associated with the $z$ integral, and the 
limits of integration in the $y'$ direction are now $\pm R'$ relative to the center of the vignetted
effective slit. Equation \ref{eq:vignetted-sinc} differs subtly but importantly from \ref{eq:evaluate-sinc} in that
$R'$ is $\theta$-dependent.  Expanding $R'$ and defining $\Delta \theta \equiv \theta - \theta_0$, 
\begin{equation}
    E_{Pv} = \frac{\phi\psi E_0 R\sqrt{\lambda}}{\pi \theta D} \sin\left( \frac{\pi D F}{\lambda} \left\{\Delta\theta^2 + \theta_0\Delta\theta \right\}\right)\,,
    \label{eq:vignetted-delta-theta}
\end{equation}
with corresponding intensity 
\begin{equation}
    I_{Pv} = I_0 \frac{R^2\lambda}{\pi^2 \theta^2 D^2 F^2}\sin^2\left( \frac{\pi D}{\lambda}\left\{\Delta\theta^2 + \theta_0\Delta\theta\right\}\right)\,.
    \label{eq:vignetted-intensity}
\end{equation}
The argument of $\sin^2$ is quadratic, rather than linear, in $\Delta \theta$; this has the effect of
changing the oscillation frequency of the sinc function near the occulter.  In particular, zeroes 
occur for every $n\in \mathbb{Z}^+$ where
\begin{equation}
    \Delta\theta^2 + \theta_0 \Delta\theta - n\frac{\lambda}{D} = 0\,,
    \label{eq:vignetted-zeroes}
\end{equation}
or 
\begin{equation}
    \Delta\theta = \frac{\theta_0}{2} \left(\sqrt{1 + 4n\lambda/D\theta_0^2} - 1 \right)\,,
    \label{eq:quadratic-formula}
\end{equation}
where the negative branch of the square root is unphysical.  If $n$ is small and the bend angle $\theta_0$ is large compared to $\lambda/D$, 
then $n\lambda/D\theta_0^2 \ll 1$;
in that case Equation \ref{eq:quadratic-formula} reduces to 
\begin{equation}
    \Delta\theta = \frac{n\lambda}{D\theta_0}\,,
    \label{eq:fraunhofer-spacing-big-theta0}
\end{equation}
so that individual dark fringes in the large-bend-angle regime follow the conventional Fraunhofer
diffraction formula and are found at $\Delta\theta$ angles of
\begin{equation}
    \Delta\theta_n - \Delta\theta_{n-1} = \frac{\lambda}{D\theta_0}\,.
\end{equation}
Contrariwise, if $n$ is sufficiently large or the bend angle $\theta_0$ is small, then $n\lambda/D\theta_0^2 \gg 1$ and in this small-bend-angle regime, Equation \ref{eq:quadratic-formula} reduces instead to
\begin{equation}
    \Delta\theta = \sqrt{\frac{n\lambda}{D}}\,,
    \label{eq:sqrt-n-step}
\end{equation}
so that individual fringes have width that varies as
\begin{equation}
    \Delta\theta_n - \Delta\theta_{n-1} \approx \sqrt{\lambda/Dn}\,
    \label{eq:fraunhofer-spacing-little-theta0}
\end{equation}
until the unvignetted condition is reached. 
Which of Equation \ref{eq:fraunhofer-spacing-big-theta0} or 
\ref{eq:fraunhofer-spacing-little-theta0} is important to the observed
pattern near an occulter in a real single-occulter instrument depends 
on the interplay between $\theta_0$, $D$, and $\lambda$, but in general if 
$\theta_0$ is comparable to or smaller than the conventional diffraction limit
of the optics, then Equation \ref{eq:fraunhofer-spacing-little-theta0} applies.  In that
case, the fringe
pattern depends only on $D$ and (weakly) on $\lambda$, not on the 
specific geometry of the occulter itself.

In Section \ref{sec:multiple}, we discuss multiple-edge occulters.  
In a multiple-edge system, the important scattering angle $\theta_0$ can be
quite small at the last stage of occultation, and 
Equation \ref{eq:fraunhofer-spacing-little-theta0} therefore drives
the appearance of the occulter edge.  For example, in a spaceborne 
coronagraph
operating at $\lambda$=650~nm, with D=250~mm, and a final scattering angle
of $\theta_0$=1.25$\arcmin$, $\lambda/D=2.6\times10^{-6}$ and 
$\theta_0^2=1.4\times10^{-7}$, so Equation 
\ref{eq:fraunhofer-spacing-little-theta0} applies, and the innermost
``bright ring'' fringe around the occulter is approximately
$5.1\arcmin$ wide.  Because the width of the fringe depends only weakly on
$\lambda$, varying the wavelength over a wide range of 450-650~nm would
 incur only a $20\%$ shift in the width of the first few fringes, which will therefore
appear nearly achromatic.

Figure \ref{fig:fringing} shows four 
examples from real instruments of the quasi-achromatic fringing effect described by 
Equation \ref{eq:fraunhofer-spacing-little-theta0}.  SOHO/LASCO-C2 uses a hybrid 
occultation scheme and therefore the innermost bright fringe is obscured by the 
internal occulter and only higher order fringes are visible at the inner occulted
edge; it ``should be'' surprising that any fringes are visible at all, given
that LASCO C2 is a broadband white-light instrument \citep{brueckner_etal_1995}.
The other three images, from multiple test-article occulters, highlight the characteristic 
extra-wide first fringe and more rapid subsequent fringes described by 
Equation \ref{eq:fraunhofer-spacing-little-theta0}. 

Equation \ref{eq:fraunhofer-spacing-little-theta0} thus explains the 
quasi-achromatic fringes seen with existing externally occulted coronagraphs.  The spacing 
of the fringes is seen to be approximately independent of the occulter
geometry itself, and to depend primarily on the length of the 
instrument between the occulter and imaging aperture. The fringe
spacing depends only weakly on wavelength, allowing even broadband
instruments such as as LASCO/C2 \citep{brueckner_etal_1995} or CATEcor \citep{deforest2024catecor}, with $\lambda / \Delta \lambda < 3$, to show visible fringes up to $n=6$ or more, 
when a na\"ive consideration might expect $n=3$ at most.

\section{Multi-Edge Scattering \label{sec:multiple}}

Section \ref{sec:half-plane} described the case of 2-D scattering around a 
single edge.  In practice, modern occulters use a multi-edge 
structure, and many multi-disk geometries have been tried, including two- and three-disk occulters \citep{brueckner_etal_1995}, 
conical multi-disk occulters \citep{brueckner_etal_1995,landini_etal_2012,yang_etal_2018}, and even smooth occulters 
as a limiting case of ``infinite'' disk count \citep{buffington2000,bout_etal_2000}.  
We note that, because the attenuation drops fastest at low values of $\theta_h$
(as seen in Figure \ref{fig:intensity-curve}), many edges (with small angular
offset each) attenuate light faster than fewer edges (with a larger angular offset each).  This 
effect was described by \citet{newkirk_bohlin_1963}, and exploited in the three-disk 
occulter of the ``C3'' and 160-disk conical occulter of the ``C2'' instruments in the Large Angle Spectroscopic COronagraph (LASCO) instrument \citep{brueckner_etal_1995}. It  
has since become a common element of occulter design in subsequent spaceborne 
instruments \citep{howard_etal_2008,yang_etal_2018,deforest_etal_2022,dudley2023}.

\begin{figure}
    \centering
    \includegraphics{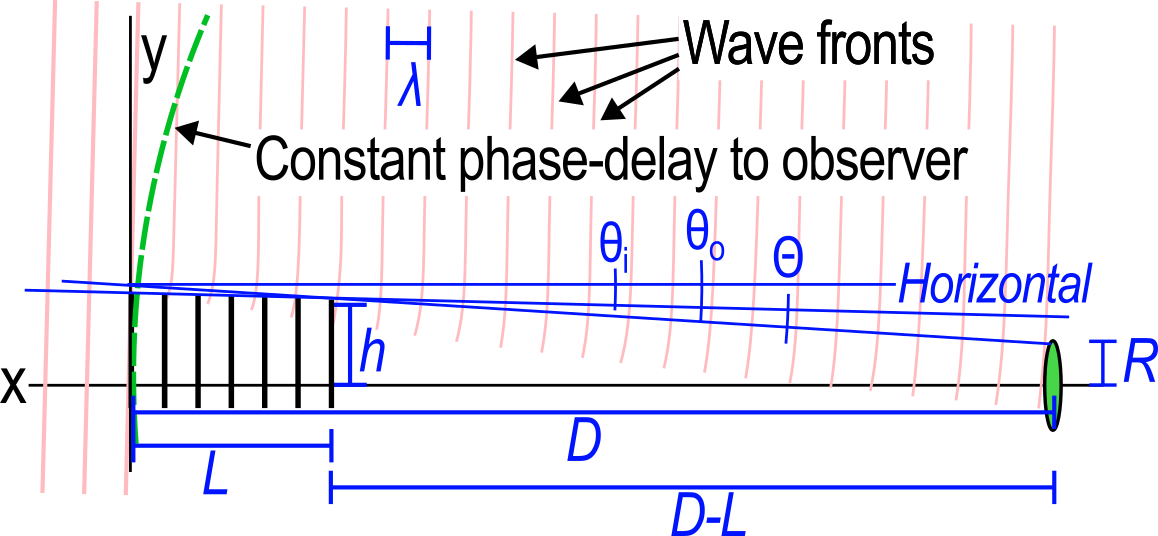}
    \caption{A basic multi-edge occulter geometry in 2-D is applicable to 
    linear heliospheric imagers (via extrusion along the $z$ axis, out of the page) or externally occulted coronagraphs (via revolution about the $x$ axis). Light incident from $\theta_i$ above the horizontal is scattered off multiple edges along an occulter of length $L$, exiting at $\theta_o$ to barely graze the edge of an optical entrance aperture.  The occulter requires the light to bend by at least $\Theta$ to enter the aperture.}
    \label{fig:multi-edge}
\end{figure}

In the general case, calculating the scattering around multiple occulter edges
is complicated enough to violate the Fresnel diffraction conditions, requiring direct
evaluation of the more complicated Kirchoff integral.  In practice, a
straightforward ``rough-and-ready'' approach is sufficient to develop intuition and 
establish a working design.

The successive-plane-wave (SPW) approximation is such an approach: Fresnel diffracted
light from each edge in a multi-edge occulter is considered to approximate a plane
wave when it encounters the next edge.
This is a reasonable approximation, because in practice multi-edge occulters
are in the regime described by Equation 
\ref{eq:fraunhofer-spacing-little-theta0} and therefore the
illuminating beam for a given edge arises from a space slightly above the prior
edge and nearly parallel to the line (in the 2-D treatment) formed by the 
tops of the two edges.  SPW puts a name and formal framework to an approximation that has already been used, without strong justification in the literature, for multi-vane occulters in circular and linear geometry \citep[e.g.,][]{buffington_etal_1998,thernisien_etal_2018}.

The SPW treatment immediately enables a simple optimization of occulter shape
design, by treating the overall multi-edge occulter as an independent
sequence of individual scatters, each of which follows Equation \ref{eq:intensity-by-angle} for a small bend angle $\Delta\Theta$ and 
radius equal to the inter-disk spacing.  We
immediately observe that, because the single-edge attenuation
is nonlinear, becoming 
more shallow with
angle (e.g., Figure \ref{fig:intensity-curve}), dividing offset angle 
equally between all active edges in a multi-edge occulter optimizes the 
overall effectiveness.  

Equipartition of angle across edges is optimal because, given two 
equally-spaced 
vanes $a$ and $b$ with different $\Delta \Theta_a$ and $\Delta \Theta_b$,
and having individual attenuation coefficients $A_a$ and $A_b$ as given by 
Equation \ref{eq:intensity-by-angle}, if (without loss of generality) 
$\Delta \Theta_a < \Delta \Theta_b$, then (from Equation \ref{eq:intensity-by-angle} and as illustrated in Figure \ref{fig:intensity-curve}), 
$-d(\textrm{log}A_a)/d\Theta_a > -d(\textrm{log}A_b)/d\Theta_b$, and 
it is beneficial to grow $\Theta_a$ at
the expense of $\Theta_b$, holding the sum constant, in the sense that
doing so improves the overall attenuation of both acting together.
Therefore, a valid way to design a multi-edge occulter 
for a particular bend angle
$\Theta$ is to start with a single-edge design, then expand the occulter
into a collection, with overall length $L$, of $n$ edges spaced equally
along the
beam direction, $\Delta L \equiv L/n$ apart and having bend
angle $\Delta\Theta \equiv \Theta/n$ each, as in Figure 
\ref{fig:multi-edge}.  Designs created this way have
edges aligned along circular envelopes in 2-D, or either cylindrical or
ogive\footnote{Readers are reminded that an \textit{ogive} is a figure of
revolution of a circle or section of a circle, about a non-diameter chord
of that circle.  While a cone or frustum has zero curvature along the 
axial direction,
an ogive has constant curvature along the axial direction.}
envelopes in 3-D for heliospheric imagers or coronagraphs, respectively.
They therefore provide more engagement of each individual vane with the propagating wavefronts,
than do conical occulters \citep[e.g.,][]{bout_etal_2000,landini_etal_2012,baccani_etal_2016,aime2020}.

The same 2-D analysis is also applicable to toroidal corral baffles such as those 
developed
by \citet{buffington2000}, as toroidal corrals are formed by revolution around a line parallel to the $y$ axis and displaced in the $+x$ direction, in Figure \ref{fig:multi-edge}.

In the SPW treatment, Equation \ref{eq:intensity-by-angle} becomes
\begin{equation}
I_{obs} = I_0 \left[\mathcal{M}\left(\Delta\Theta\sqrt{\pi\Delta L/\lambda}\right)\right]^{n-1}\mathcal{M}\left(\Delta\Theta\sqrt{\pi(D-L)/\lambda}\right)\,,
    \label{eq:multiscatter-intensity}
\end{equation}
where the auxiliary function $\mathcal{M}$ is just the complicated part of Equation \ref{eq:intensity-by-angle}:
\begin{equation}
    \mathcal{M}\left(\gamma\right) \equiv \frac{1}{\pi}\left\{\left[\sqrt{\pi/8} - \mathcal{C}\left(\gamma\right)\right]^2+\left[\sqrt{\pi/8}-\mathcal{S}\left(\gamma\right)\right]^2\right\}\,.
    \label{eq:define-mathcal-M}
\end{equation}
The first $n-1$ coefficients in Equation \ref{eq:multiscatter-intensity}
represent scatters by the first $n-1$ disks; the last term is broken out
because the scatter is over the generally-much-longer distance all the way 
to the focusing optics.  In Equation \ref{eq:define-mathcal-M}, $\mathcal{M}$ takes 
the form of an analytic propagator for edge scattering, and is a simpler form than
the propagator derived by \citet{wang_etal_2021} through more rigorous analysis.

Equation \ref{eq:multiscatter-intensity} is computable but complicated.
More simplification is possible, to reveal the essential behavior of the 
propagator.  For a typical solar occulter, 
$\Delta\Theta$ might be on the order of 1-2~arcmin, and $\Delta L$ might be 
on the order of 5~mm.  For those numbers and a wavelength of 650~nm, the argument to $\mathcal{M}$ evaluates to 0.05--0.1.  That affords the approximations
\begin{equation}
    \mathcal{C(\gamma)} \approx \gamma
    \label{eq:small-C}
\end{equation}
and
\begin{equation}
    \mathcal{S(\gamma)} \approx 0\,,
    \label{eq:small-S}
\end{equation}
which may be verified from their definitions in Equations \ref{eq:C-function} and \ref{eq:S-function} when $\gamma$ is small.  In turn, that yields
\begin{equation}
    \mathcal{M}\left(\gamma\right) \approx \frac{1}{\pi}\left\{\left[\sqrt{\pi/8} - \gamma\right]^2 + \pi/8\right\} \approx \frac{1}{\pi}\left\{\frac{\pi}{4}-2\gamma\right\} = \frac{1}{4}\left[1 -\frac{8\gamma}{\pi}\right]
    \label{eq:approximate-mathcal-M}
\end{equation}
for those coefficients.  
Equation \ref{eq:multiscatter-intensity} thus simplifies to
\begin{equation}
    I_{obs} \approx I_0\left[\frac{1}{4}\left(1 - 8\Delta\Theta\sqrt{\frac{\Delta L}{\pi\lambda}}\right)
    \right]^{n-1}   
    \mathcal{M}\left(\Delta\Theta\sqrt{\frac{\pi(D-L)}{\lambda}}\right)\,.
    \label{eq:multiscatter-approximate-I}
\end{equation}

Equation \ref{eq:multiscatter-approximate-I} points intuitively to the reason why conical coronagraph 
occulters \new{\citep[e.g.,][]{bout_etal_2000,landini_etal_2012,Landini_etal_2017}} work.  In the ray approximation, a conical multi-disk occulter
``should'' yield no improvement over a dual-disk occulter, because the intermediate disks
don't intersect any rays.  But the intermediate disks do interact with wave fronts 
and further
attenuate stray light that, in the ray approximation, would merely graze across the top of
each disk in the conical assembly.  So long as the SPW approximation is valid, this yields
a minimum attenuation of a factor of $4$ per disk (or vane in a linear occulter).

In a circular-profile design, each disk attenuates slightly more than a factor of 4, because of the 
perturbation coefficient in parentheses, inside the square brackets of Equation 
\ref{eq:multiscatter-approximate-I}.  In the specific case of $n=10$, $\Delta L=5$mm, $\Delta\Theta = 1.5\arcmin$, 
and $\lambda=650$nm, the 
perturbation coefficient evaluates to 0.83, improving the 
total attenuation by nearly an order of magnitude.  But as $n$ increases (at constant 
$L$ and $\Theta$), both 
$\Delta\Theta$ and $\Delta L$ decrease, and the perturbation coefficient rapidly approaches unity.
In the large-$n$ case that coefficient may be neglected entirely, yielding 
\begin{equation}
    \lim_{n\rightarrow \infty} I_{obs} = I_0 (1/4)^n\,.
    \label{eq:smooth-I}
\end{equation}
This large-$n$ case has been described by several authors, with varying degrees
of analytic rigor, since the mid 20th Century \citep{newkirk_bohlin_1963}. 
Equations \ref{eq:multiscatter-approximate-I} and \ref{eq:smooth-I} represent a
middle way between intuitive observation \citep{buffington_etal_1998} and
more recent detailed analytic treaments \citep{wang_etal_2021} that reveal similar 
results through rigorous analysis followed by formal approximation.

Because Equation \ref{eq:smooth-I} predicts much better performance for 
multi-edge occulters
as $n$ increases, which is not borne out in the ``infinite-edge'' case of smooth
occulters \citep[e.g,][]{bout_etal_2000}, it is useful to explore the practical limits on $n$.  
Mechanical tolerance is one limit: 
in the same hypothetical case as above, with $n=10$, $\Delta L=5$mm, 
$\Delta\Theta=1.5\arcmin$, and $\lambda=650$nm,
the edges must have $h$ placement precision (including both fabrication 
and alignment errors)
of $\pm$2~$\mu$m to ensure all edges interact with the beam; and adding more edges (within the same length) tightens 
the positional tolerance
according to $n^{-2}$.  Indeed, this effect was thought to be the reason 
that \citet{newkirk_bohlin_1963} did not observe the theoretical benefit when they 
fabricated, assembled, and tested a 140-disk occulter made from individual
graduated disks on a spindle.

But, surprisingly, mechanical tolerance is not necessarily the limiting factor for $n$: the SPW approximation itself fails at high $n$. Smooth
occulters, which represent the $n\rightarrow \infty$ limit, are not 
infinitely effective.
To find the limits of SPW, we explore its breakdown in two different 
ways:  via a hybrid
ray-diffraction approach, and via analysis of how wavefronts at each 
edge change from
simple plane-waves (or even constant-radius wave fronts) as $n$ increases.

\begin{figure}
    \centering
    \includegraphics[width=3in]{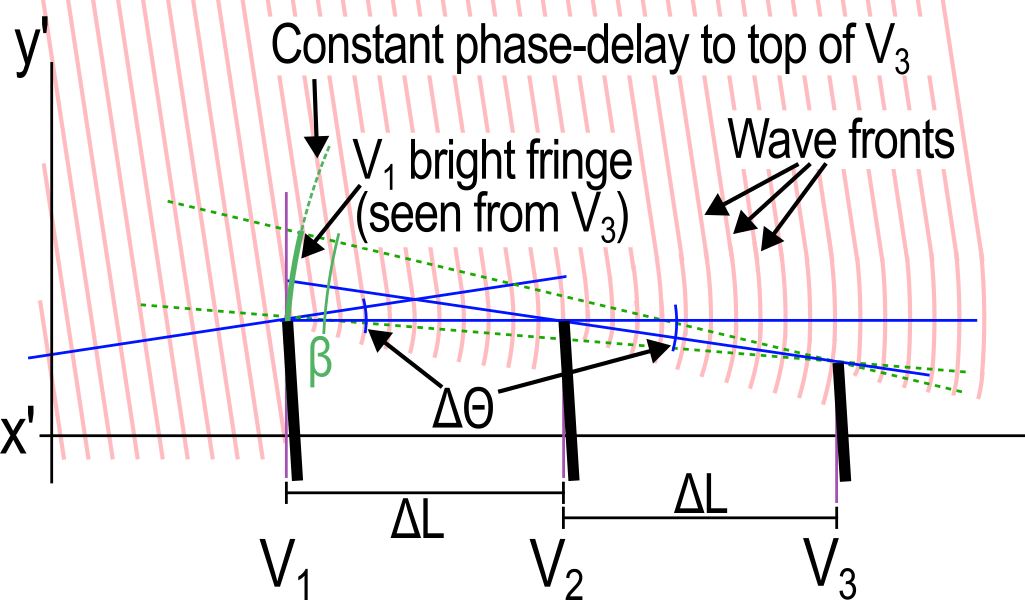}
    \caption{Close-up rendering of the geometry in Figure \ref{fig:multi-edge} illustrates a 
    limit of the successive plane wave (SPW) approximation, using a hybrid ray approach. As $\Delta L$ and/or $\Delta \Theta$ shrink, $V_2$ fails to fully obscure rays from the first bright fringe above $V_1$, as seen from the top of $V_3$.  Note the tilted
    ($x'$,$y'$) frame of reference, selected to make $V_1$--$V_2$ horizontal on the figure.}
    \label{fig:multi-edge-limit}
\end{figure}

\subsection{Limits of the SPW approximation: hybrid approach\label{sec:SPW-breakdown-hybrid}}
Figure \ref{fig:multi-edge-limit} is a close-up 
rendering of a portion
of the occulter in Figure \ref{fig:multi-edge}, showing two approaches to understanding 
the breakdown of SPW. Three vanes, separated by $\Delta L$ 
in space and $\Delta\Theta$ in
angle, form part of an occulter with cylindrical (in extruded linear geometry) or ogive (in 
revolved circular geometry) overall envelope.  For convenience, the $x$ and $y$ directions are
replaced with $x'$ and $y'$, rotated so that the $V_1$--$V_2$ direction is along the
$x'$ axis.  Incident plane waves arrive at $V_1$ rising at an angle $\Delta\Theta$ and
interact with the edge.  The $V_2$--$V_3$ direction drops by $\Delta\Theta$ relative to
$V_1$--$V_2$.  

In the ray approximation, $V_2$ in Figure \ref{fig:multi-edge-limit} is high enough to
prevent rays from $V_1$ from reaching $V_3$.  However, diffracted light that passes over $V_1$ to
reach $V_3$ actually passes mainly
through the space just above $V_1$.  That space is approximated by the locus, pictured 
in the left panel of Figure \ref{fig:multi-edge-limit} and labeled ``$V_1$ bright edge'',
where the Fresnel integrand $u$ in Equations \ref{eq:C-function} and \ref{eq:S-function} varies by roughly unity, i.e. from near 0 to less than 1. That 
approximation holds because, in the case under consideration, the $h/s$ value 
for $V_1$ itself,
relative to $V_3$, is close to zero, and the first unit of range in the $\mathcal{C}$ 
and $\mathcal{S}$ integrations contributes most strongly to the total value of the Fresnel integrals in Equation \ref{eq:intensity-by-angle}.  That
effect is related to, but not the same as, the fringing discussed in Section \ref{sec:fringes}: that fringing requires interaction between an aperture and an occulter edge, while this effect has to do with the locations on the $y'$ axis that contribute most strongly to the Fresnel integrals.
If that locus is directly visible from the top of $V_3$, the edges do not act
independently and 
the SPW approximation becomes invalid.  Thus the SPW treatment requires at least
\begin{equation}
    \beta < \Delta\Theta\,,
    \label{eq:beta}
\end{equation}
where $\beta$ is labeled in Figure \ref{fig:multi-edge-limit}, and is given by 
\begin{equation}
    \beta = \frac{h'}{2 \Delta L}
\end{equation}
and
\begin{equation}
    h' = s \equiv \sqrt{2\Delta L \lambda / \pi}\,,
\end{equation}
where $h'$ is height above $V_1$ and $s$ is calculated (using Equation \ref{eq:define-s}) for the $V_1$--$V_3$ gap 
in the absence of $V_2$.  So the overall requirement for the successive-plane-wave approximation is 
\begin{equation}
    \Delta\Theta > \frac{\sqrt{2\Delta L \lambda / \pi}}{2\Delta L} = \sqrt{\frac{\lambda}{2\pi\Delta L}}
    \label{eq:beta-2}
\end{equation}

Equation \ref{eq:beta-2} can be reduced further, by noting that in an $n$-edge 
occulter of
length $L$, $\Delta\Theta=\Theta/n$ and $\Delta L=L/n$, to arrive at a limiting $n$ for 
given occulter parameters:
\begin{equation}
    \Theta/n > \sqrt{\frac{n\lambda}{2\pi L}}
    \label{eq:beta-3}
\end{equation}
or, solving for $n$, 
\begin{equation}
    n < \left(\frac{2\pi\Theta^2L}{\lambda}\right)^{1/3}\,.
     \label{eq:n-limit}
\end{equation}
For example, for an occulter with $\Theta=0.5^\circ$ and $L=75$~mm operating 
at 650~nm, the maximum
$n$ is roughly 4.  At higher $n$, one might expect occulter performance to 
diverge from the SPW model, becoming worse because each vane no longer
obscures the bright edge of the prior vane, from the successive vane. However, SPW-based
calculations have proven to be good estimators of occulter performance in multiple
geometries even when Equation \ref{eq:n-limit} implies they should not \citep{laurent_etal_2025, Erickson_etal_2025}.  Understanding why requires
a slightly less facile treatment, which we explore in the next section.

\begin{figure}
    \centering
    \includegraphics[width=0.67\linewidth]{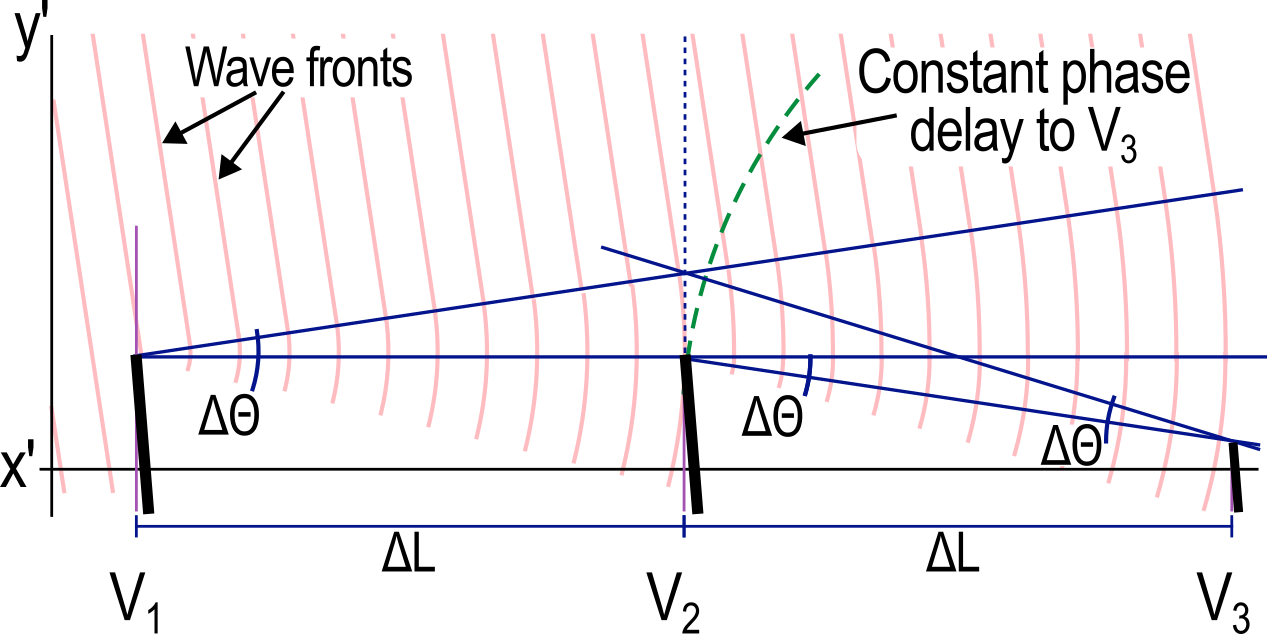}
    \caption{A second close-up rendering of the geometry in Figure \ref{fig:multi-edge}
    recaps Figure \ref{fig:multi-edge-limit} from a wavefront perspective.  Following
    Huygens' principle, perturbed wavefronts may be 
    approximated (in 2-D) by splicing a circular arc onto the unperturbed part of
    the 
    wave front.  Taking the Huygens' wavelet approach along the dotted line
    above $V_2$ thus yields two Fresnel integrals: one for the portion of 
    the wavefront that is perturbed by $V_1$ (and therefore curves with 
    radius $\Delta L$), and one for the unperturbed, straight portion of 
    each wavefront.  For convenience the incident angle onto $V_1$ is taken to be
    the same as $\Delta \Theta$ for each vane of the occulter as a whole.  The ($x'$,$y'$) axes are selected to make the $V_1 - V_2$ line horizontal, which makes the vanes themselves appear slanted.
    }
    \label{fig:multi-edge-limit-2}
\end{figure}

\subsection{Limits of the SPW approximation: spliced-wavefront approach\label{sec:SPW-breakdown-wavefront}}

Full treatment of a multi-edge system requires evaluating the complete 
Kirchoff diffraction integral \citep[][Chap. 8]{bornwolf1999}, of
which the Fresnel integral (Equation \ref{eq:fresnel-integral}) is but an 
approximation. 
Many authors discuss the use of numerical methods to evaluate that integral 
\citep[e.g.,][]{breault_1977, gong_socker_2004, yang_etal_2018}.  But it
is possible to understand the limits and form of SPW violation even without resorting
to numerical methods, by 
considering the way in which the wavefronts begin to violate the conditions for 
both the SPW and Fresnel 
approximations as $n$ increases.

Wavefront shape evolves in two important ways as $n$ increases (at constant $L$). 
The first is illustrated at the $V_2$ edge in Figure \ref{fig:multi-edge-limit-2}. The
wave fronts arriving at the $V_2$ edge from $V_1$ are not actually planar, and can 
each be better approximated with (in 2-D) a circular segment of radius $\Delta L$, 
spliced to a plane-wave segment that is perturbed only negligibly by the interaction
with $V_1$.  The SPW approximation treats this approximately circular 
segment of the wavefront at $V_2$ (and subsequent vanes) as negligible to
the overall shape of the wavefronts as they propagate along the 
occulter.

While the exact shape of the wavefront at $V_2$ is complicated to describe 
analytically, it is straightforward to approximate it by simply splicing a
circular and linear segment together.  For the case of $V_2$,
Equation \ref{eq:simpler} breaks into two integrals.  $E_3$, the electric field at 
the top of $V_3$ in Figure \ref{fig:multi-edge-limit-2}, can be written
\begin{equation}
    E_{3} = \frac{\phi}{\sqrt{\pi}} \left( 
    \int_{\epsilon}^{2 \epsilon} E_{2}(u)\, e^{i 2 u^2 } du + 
     e^{i 2 \epsilon^2} 
     \int_{2 \epsilon}^\infty E_{2}(u)\, e^{i u^2 } du \right)\,,
     \label{eq:spliced-1}
\end{equation}
where 
\begin{equation}
    \epsilon \equiv \Delta L \Delta \Theta / s = \Delta\Theta\sqrt{\frac{\Delta L \pi}{\lambda}}\,,
    \label{eq:epsilon}
\end{equation}
and the function 
$E_{2}(u)$ is the calculated electric field at each location along the line
of integration above $V_2$.  $E_{2}(u)$ may be calculated directly using Equation
\ref{eq:simpler} applied to an observer at the corresponding location above $V_2$, 
due to Fresnel diffraction effects imposed by $V_1$ and the incident plane waves.  
The initial factor in Equation \ref{eq:spliced-1} is the same as in Equation
\ref{eq:simpler}.

The first integral of Equation \ref{eq:spliced-1} describes the contribution to 
$E_3$ from space between the top 
of $V_2$ (which is 
elevated above the direct line of sight from $V_3$ to $V_1$ by the distance 
$\Delta \Theta L$) to the top of the spliced circular approximate wavefront (which is
$\Delta \Theta L$ higher still). The imaginary exponent in the integrand has a $2u^2$
term describing the doubled curvature from the two circular constant-phase surfaces:
one from $V_1$ and one to $V_3$.  It also has a linear term describing the opening angle
between the two constant-phase surfaces, which is $\Delta \Theta$ at the top of $V_2$.

The second integral of Equation \ref{eq:spliced-1} describes the contribution to $E_{3}$ 
from the space above the top of the spliced circular wavefronts, where the wavefronts
are approximated with straight lines.  The quadratic term is dependent only on $u^2$
and not on $2u^2$, accounting for the single curvature of the constant phase-delay
surface to $V_3$ from the line of integration.  The imaginary exponential term in front
of the integral matches the phase delay at the splice point between the linear and 
circular approximations.

Cleaning up the 
integration with another substitution $u=v/\sqrt{2}$,
\begin{equation}
     E_{3} = \frac{\phi}{\sqrt{\pi}} \left( 
    \frac{1}{\sqrt{2}}\int_{\sqrt{2} \epsilon}^{2\sqrt{2} \epsilon} E_{2}(v/\sqrt{2})\, e^{i v^2 } dv + 
     e^{i 2 \epsilon^2} 
     \int_{2 \epsilon}^\infty E_{2}(u)\, e^{i u^2 } du \right)\,.
     \label{eq:spliced-2}
\end{equation}

Calculating $E_{3}$ with Equation \ref{eq:spliced-1} or \ref{eq:spliced-2} is complicated in the general
case.  However, we are not interested in the general case, but in the case where 
$\Delta \Theta$ is small and where $\Delta L$ is at most comparable to $s$.  Following
the example of Fresnel himself, we can therefore simplify Equation \ref{eq:spliced-1} by
aggressively approximating and/or neglecting effects wherever possible.

In particular, $\epsilon \ll 1$ and therefore 
Equations \ref{eq:small-C} and \ref{eq:small-S} hold,
which simplifies the problem considerably.  
Further, we can approximate $E_2$ as piecewise constant in $u$.  In 
the linear regime we set $^{|}E_2(u)=E_0$, and in the circular regime, we allow
$^{\circ}E_2$ to attenuated by the constant attenuation term for Fresnel diffraction around $V_1$ onto the edge of $V_2$ (where the $|$ and $\circ$ mark which regime is being used for the calculation).  Applying the relevant geometry to Equation \ref{eq:e-field}, and applying Equations \ref{eq:small-C} and \ref{eq:small-S}, yields
\begin{equation}
    ^{\circ}E_2 = \phi' E_0\left\{\frac{1}{2} - \frac{1}{\sqrt{\pi}}\epsilon\right\}\,,
    \label{eq:E2}
\end{equation}
where $\phi'$ is a phase factor approximately equal to $e^{i\pi/4}$.

Substituting for $^{|}E_2$ and $^{\circ}E_2$, neglecting two small phase shifts, and using Equations \ref{eq:small-C} and \ref{eq:small-S} both evaluates the integrals from Equation \ref{eq:spliced-2} and simplifies the remaining terms:
\begin{equation}
    E_3 = \frac{\phi''E_0}{\sqrt{\pi}}
    \left( 
    \left(  \frac{1}{2}-\frac{\epsilon}{\sqrt{\pi}}\right)  \epsilon  + 
    \sqrt{\pi} \left( \frac{1}{2} - \frac{\sqrt{2}}{\sqrt{\pi}}\epsilon \right)
    \right)\,,\label{eq:spliced-3}
\end{equation}
where $\phi''$ rolls up the phase shifts.  Equation \ref{eq:spliced-3} simplifies further, to  
\begin{equation}
    E_3 = \phi''E_0
    \left( 
    \frac{1}{2} - \frac{1+\sqrt{2\pi}}{\pi}\epsilon - \frac{1}{\pi}\epsilon^2
    \right)\,.\label{eq:spliced-4}
\end{equation}
Equation \ref{eq:spliced-4} yields a separate estimate, independent of Equation 
\ref{eq:beta-3}, of when and how the SPW model breaks down at low values of 
$\Delta L$ and/or $\Delta \Theta$.  In particular, SPW predicts 
(Equation \ref{eq:approximate-mathcal-M}) that the combined effect of $V_1$ and
$V_2$ on the illumination at $V_3$ should be an overall attenuation of better than 
$1/4$ in E (or, equivalently, $1/16$ in I).  That relation only holds if 
\begin{equation}
    \epsilon > 0.22...\,.
\end{equation}
Expanding $\epsilon$ (via Equation \ref{eq:epsilon}) and solving for $n$ as 
in the derivation of Equation \ref{eq:n-limit},
\begin{equation}
    n < \left( \frac{20\pi\Theta^2 L}{\lambda} \right)^{1/3}\,,
    \label{eq:n-limit-2}
\end{equation}
which is considerably less stringent than Equation \ref{eq:n-limit}.  For the same 
occulter parameters as considered in Section \ref{sec:SPW-breakdown-hybrid}, the
spliced wavefront approach predicts agreement with SPW for occulters with at least 8 
vanes.

Examining the wavefronts at right in Figure \ref{fig:multi-edge-limit-2} reveals
that, at high $n$, the two-wavefront splicing approach that led to Equations 
\ref{eq:spliced-1} -- \ref{eq:spliced-4} is pessimistic about the overall 
attenuation of multi-vane occulters. There are two reasons for that pessimism.

Firstly, in calculating $E_3$ we treated the strength of the linear section of the
wave fronts as equal to the incident field strength, but this is not the case.  In 
particular, the nomogram in Figure \ref{fig:cornu-spiral} and the demonstration 
attenuation plots in Figure \ref{fig:intensity-curve} show that the incident 
wavefronts at $V_2$ are attenuated significantly at small negative values 
of $\epsilon$ (or equivalently, $h$), i.e. in exactly the portion of the linear wavefront segment 
that contributes most strongly to the intensity at $V_3$.  The approximation makes
the integrals in Equation \ref{eq:spliced-2} far more tractable by forcing them into
the same form as in Equation \ref{eq:simpler}, but also makes Equation 
\ref{eq:n-limit-2} pessimistic by a factor of order $2^{1/3}$.  

Secondly, wavefronts near successive
vanes are further modified by each prior vane, beyond the simple two-segment splice
we described.  In Figure \ref{fig:multi-edge-limit-2}, the wavefront over 
$V_3$ is better described by three
segments with piecewise-constant curvature: the lowest portion is formed by the 
last few Huygens' wavelets just above $V_2$, and has radius $\Delta L$.  The middle
portion is formed around $V_1$ and has radius
$2\Delta L$.  As the wavefronts propagate along the occulter to subsequent vanes,
more splices are required; and the effect is to cause occulters with large $n$ to 
attenuate slightly more than the subsequent simple splices suggest.  

These two effects suggest that the regime of SPW applicability is best approximated
with the formula
\begin{equation}
    n_{max} \sim \left( \frac{50\pi\Theta^2L}{\lambda}\right)^{1/3}\,,
    \label{eq:n_max}
\end{equation}
where the factor of 50 arises from a combination of the $2^{1/3}$ from the $^|E$ attenuation and a small correction for the multiple-radius curve in $^\circ E$. 
Equation \ref{eq:n_max} implies a practical maximum of 11-12 vanes for the example occulter described
above, before performance begins to fade below the SPW prediction, and further 
explains why smooth occulters perform only marginally, and not infinitely, better
than multi-vane occulters even under ideal conditions.  

\section{Disk Occulters\label{sec:disks}}

Coronagraph external occulters are typically circular.  Revolving the 
system 
of Figure \ref{fig:fresnel-geometry} about the $x$ axis yields a typical 
coronagraph geometry, with one or a series of disk edges blocking sunlight 
from entering the instrument.  In general, the geometry is set such that the
final exit angle $\theta_0$ from the occulter (visible in Figure
\ref{fig:multi-edge}) becomes the cone angle of a ``shadow frustum'' whose
large end, at the rear of the occulter, has radius $h$ and whose small end,
at the entrance aperture, has radius $R$, slightly larger than the entrance
aperture of the optics.  $\theta_i$ is tuned to be slightly larger than the
apparent size of the Sun.

Many aspects of coronagraph design are 
agnostic to whether the occulters of Figures 
\ref{fig:fresnel-geometry},\ref{fig:fringe-geometry}, 
and \ref{fig:multi-edge} are extruded along the third dimension 
(as in a linear-geometry heliospheric imager) or revolved about the $x$ axis
(as in a circular-geometry coronagraph), with two important exceptions. 

The first important feature of revolved (cylindrical, conic, or ogive)
occulters is in the angular image domain: brightness of the fringes discussed
in Section \ref{sec:fringes} scales
inversely as the apparent distance $\varepsilon$ from the boresight of
the instrument.  In this
geometry, Equation \ref{eq:I_Pnv-avg} becomes
\begin{equation}
    ^\circ\bar{I}_{Pnv} = I_0 \frac{R^2 \lambda}{2 \pi^2 F^2 (\varepsilon-\varepsilon_{min})^2\varepsilon D^2 }\,,
    \label{eq:I_Pnv-avg-circ}
\end{equation}
where $^\circ\bar{I}_{Pnv}$ is the average brightness of the fringes in the
circular geometry, $\varepsilon_{min}$ is the apparent size of the
inner 
field-of-view cutoff of the detector, and (as in Equation \ref{eq:I_Pnv-avg}) $^\circ\bar{I}_{Pnv}$ is calculated as a deposited power per unit
area at the focal plane and thus depends on the focal length of the optics, $F$.  The additional factor of 
$\varepsilon^{-1}$ arises because, as apparent distance from the centerline
increases, the light from a given small length of occulter edge is spread
across more of the focal plane. Furthermore, far from the apparent edge 
of the 
occulter, $\varepsilon_{min}$ is negligible and $^\circ\bar{I}_{Pnv}$ 
drops off as $\varepsilon^{-3}$; this form matches the
known dropoff rate of the coronal brightness \citep[e.g.,][]{DeForest_etal_2016}, ensuring that
diffracted stray light does not dominate the far field.  \new{The circular geometry enables
several other approaches to the diffraction calculation, such as using Babinet's Principle to 
replace the definite integral in Equation \ref{eq:fresnel-integral} with one over the complementary
finite domain.}

The second important exception is that, in the spatial domain, coherence of incoming rays around the occulter leads to constructive interference and a non-intuitive ``Arago spot'' 
\citep{arago1819} along the centerline of the instrument. Famously, the spot from a point source and circular disk in ideal geometry forms 
a $J_0$ Bessel function with peak value equal to the intensity of the 
incident light \citep[e.g.,][]{bornwolf1999,reisinger_2017}. 

For small beam deviations from the centerline, the Arago spot remains
nearly unchanged, and a simple occulting disk therefore forms a real
image of a distributed incident light source, convolved with a Bessel function
spot profile on-axis and a compact elliptical evolute form
off-axis \citep{coulson_1922,Harvey1984}.
This is the basis of the field of Fresnel zone-plate imaging
\citep[e.g.,][]{hecht_zajac_1974, davila_etal_2011}. Single-disk 
occulters therefore also act as pinhole cameras, forming a real image
of the Sun at the center of the occulted shadow.
But zone-plate imaging isn't relevant to circular multi-edge occulters.  
The solar image
is destroyed 
by the multiple scattering events, at least in the SPW approximation used to derive 
Equation \ref{eq:multiscatter-intensity}. Each disk in the occulter scatters 
slightly converging rays from the prior disk, eliminating whatever image 
information may in principle be present after the first scattering event.  
The central bright spot from a multi-disk occulter is therefore just 
that -- a single spot with an approximate Bessel function profile.

\begin{figure}
    \centering
    \includegraphics[width=0.5\linewidth]{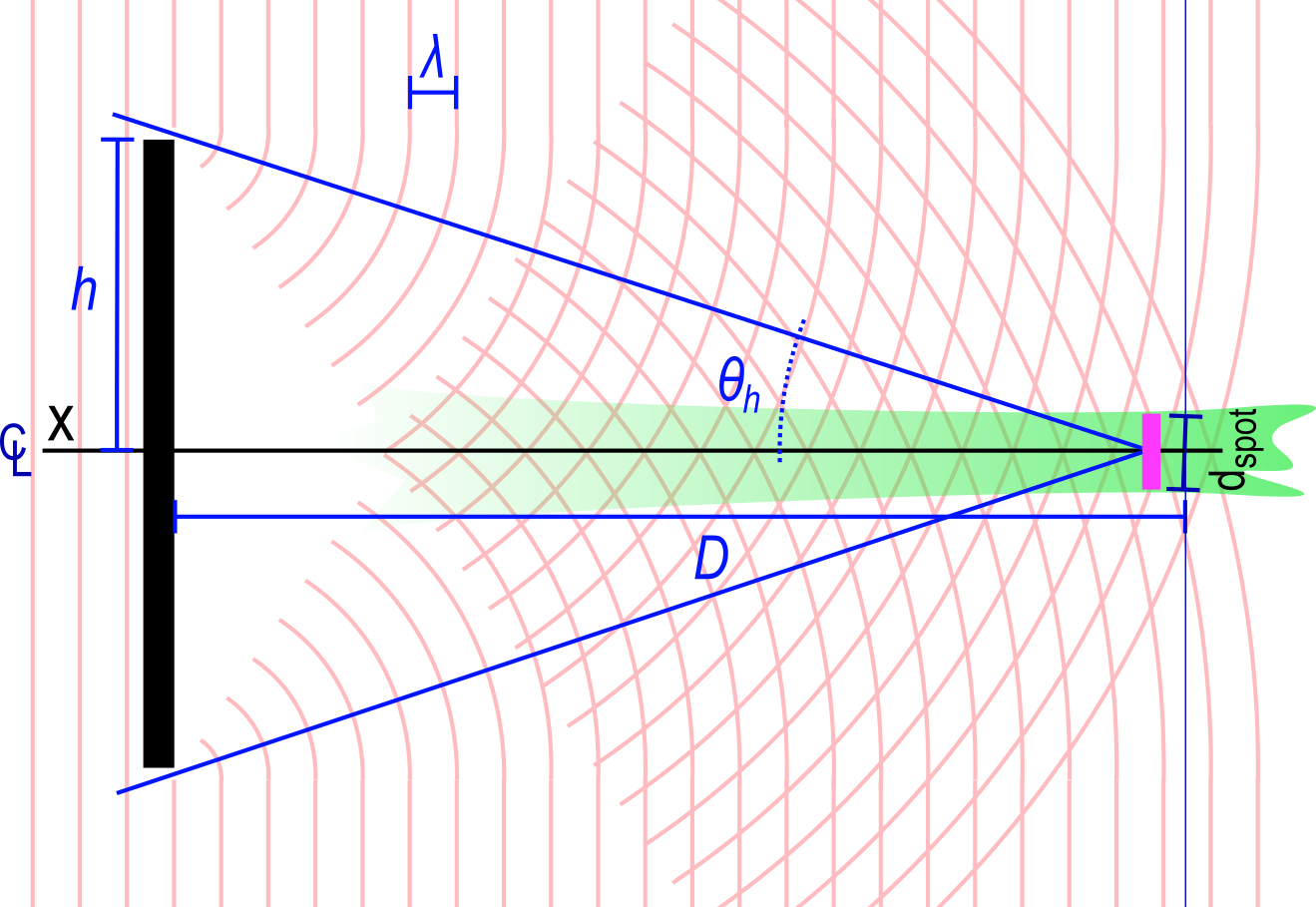}
    \caption{The Arago spot is a counterintuitive optical phenomenon that arises from interference around the periphery of a circular occulter.  The occulter from Figure \ref{fig:fresnel-geometry} is revolved about the X axis.  Huygens' wavelets from around the entire edge of the occulter interfere constructively at a small locus along the centerline of the shadow.  The width of this bright spot is determined by
    Fraunhofer-style interference between the contributions from opposite sides of the occulter.}
    \label{fig:arago-spot}
\end{figure}

The Arago spot size may be \new{estimated} directly using a variation of 
Fraunhofer interference.  
In Section \ref{sec:fringes}, we discussed the formation of
directional fringes in the Fresnel-diffracted light around an edge.  In the case
of the Arago spot, light from opposite sides of the circular occulter interferes 
constructively at the central axis as in Figure \ref{fig:arago-spot}. Under
small offsets, the two beams go out of phase.  Opposite sides of the ``bright ring'' illuminating
the point of interest go fully out of phase when $h\phi=\lambda/4$.  The ring as a whole 
exhibits full cancellation when $h\phi=\lambda/2\sqrt{2}$, with the additional $\sqrt{2}$ arising
from the amount of total edge length at each vertical offset around the perimeter of the occulter.
So the Arago spot behind a multi-disk occulter has a size given by:
\begin{equation}
    r_A = \phi_0 \left(D - L\right) \approx \frac{\lambda \left(D - L\right)}{2\sqrt{2} h_n }\,,
    \label{eq:spot-size}
\end{equation}
where $r_A$ is the radius of the spot, $\phi_0$ is the radius of the edge of the spot, $D-L$ is the
distance from the final disk to the location where the spot is measured, and $h_n$ is the radius of
of the final disk in the multi-disk occulter.  

As an example, an instrument operating at $\lambda=650$~nm, with
$D-L=175$~mm, and $h=10$~mm would have an Arago spot diameter
of 4~$\mu$m at the aperture.  A single-disk device operating on a 
10~mm aperture would thus have a minimum attenuation coefficient of 
order $10^{-7}$ from its Arago spot alone.  Occulters with $n$ disks, 
operating
in the SPW regime, generate Arago spots with intensity that is 
attenuated by the 
effect of the first $n-1$ disks, and the residual Arago spot is 
therefore usually negligible in its overall contribution.

Far from
the Arago spot, interference effects may be neglected and 
incident diffracted beams on different parts of the aperture may be treated
as incoherent.  Considering a single 2-D cross-section of a circular
aperture
(as pictured in Figure \ref{fig:multi-edge}), one may consider $I_{obs}$ at the aperture to
be a function of $y$.  Varying $y$
only affects the final attenuation from bending around the final ($n^{th}$) disk and not the geometry nor attenuation from disks 1 through $n-1$; thus,
\begin{equation}
    I_{obs}(y) = I_0 \mathcal{A}_{f} \mathcal{M}\left(\left[\Delta\Theta + \frac{R-y}{D-L}\right]\sqrt{\frac{\pi(D-L)}{\lambda}}\right)\,,
    \label{eq:local-aperture-brightness}
\end{equation}
where $\mathcal{A}_{f}$ is the overall attenuation from the first $n-1$ disks of the occulter.  Migrating from 2-D to 3-D by revolving around
the $x$ axis, the average scattered intensity across the
aperture is the integral of contributions from each $y$ value, divided by the total area of the aperture.  Thus,
\begin{equation}
    I_{ap} = \frac{I_0 h \mathcal{A}_{f}}{R}\int_{-R}^R
    \mathcal{M}\left(\left[\Delta\Theta+\frac{R-r}{D-L}\right]\sqrt{\frac{\pi(D-L)}{\lambda}}\right)\,dr\,,
    \label{eq:average-aperture-brightness}
\end{equation}
where the 
the variable of integration is $r$ instead of $y$ to reflect that the 2-D 
system of Equation \ref{eq:local-aperture-brightness} has been revolved
around the $x$ axis for a circular occulter and aperture. Two 
irregularities stand out: Equation \ref{eq:average-aperture-brightness} has 
an $R$ in the denominator and lacks an expected 
$\pi r$ in the integrand.  
The denominator $R$ arises because the average covers twice the area of the aperture, 
i.e. $2 \pi R^2$.  This
cancels
with a factor of $2\pi R$ in the numerator to yield the $R$.  The factor of $2\pi R$ and the lack
of a $2\pi r$ under the integral both arise because light 
scattered into each portion of the aperture is treated as the incoherent
sum of contributions of light from all around the aperture, and stray light
is therefore concentrated near the center of the aperture.  This 
yields a factor of $R/r$ in the integrand, canceling out the expected $r$; 
and the remaining $2\pi R$ is constant and migrates outside the integral.
The concentration  effect is similar to, but distinct from, the Arago spot
itself: the 
Arago spot is formed by the \textit{coherent} concentration of electric
field from all around the perimeter of the occulter; while this far
less pronounced
effect is due to the \textit{incoherent} concentration of 
visible flux from all around the perimeter, and merely equalizes the 
contribution of each radius across the full aperture.

Because $\mathcal{M}$ varies rapidly with its angular argument over the 
long throw between the occulter and the aperture, the aperture-average 
scattered-light intensity 
in Equation \ref{eq:average-aperture-brightness} is dominated by the outermost (high-$r$) portion
of the 
aperture. The sharpness of this bright outer boundary in the spatial domain
is determined by the specific geometry of the instrument.  Fortunately, Equation 
\ref{eq:average-aperture-brightness} is straightforward to evaluate
numerically for any given geometry.

For initial occulter design, where precision is less important than 
rapid evaluation, one may neglect the integral in Equation 
\ref{eq:average-aperture-brightness} entirely, and consider all of the
final disk's diffracted light to land on the outermost edge of the
aperture.  This simplification yields
\begin{equation}
    I_{ap} \approx \frac{I_0hA_f}{R}\ \mathcal{M}\left(\Delta\Theta\sqrt{\frac{\pi(D-L)}{\lambda}}\right)\,,
    \label{eq:approx-aperture-brightness}
\end{equation}
which is really just
\begin{equation}
    I_{ap} \approx I_{obs} \frac{h}{R}
    \label{eq:approx-aperture-brightness-2}
\end{equation}
where $I_{obs}$ is taken from Equation \ref{eq:multiscatter-approximate-I}.
In other words, to a reasonable and somewhat conservative approximation, all 
of the scattered light around 
the occulter is focused onto a ring around the outside of the aperture
(neglecting the Arago spot, which is very bright but also very small), and
the average diffracted intensity across the aperture may be approximated 
by scaling the basic SPW calculation for a simple multi-vane barrier in 2-D.

Solar disk occulters in general are not obstructing a point source, but
rather the Sun itself.  Therefore, 
Equation \ref{eq:multiscatter-approximate-I} only applies if the apparent
size of the Sun is small compared to the apparent size of the occulter
as seen from the instrument focusing optics.  \new{This is the case for 
typical linear geometries (heliospheric imagers), but not for typical
coronagraph geometries.}  Fortunately, the problem
of a distributed source is exactly analogous to the problem of a 
distributed aperture, and the same design approximation may be used
on the input side of the occulter as the output side. \new{The same
process may be followed for the disk of the Sun, as for the aperture. This
yields an equation similar to 
Equation \ref{eq:average-aperture-brightness}, but with the integral
running over entrance angle rather than aperture location; and for design
purposes it is a helpful and conservative assumption to treat the Sun as a
bright ring.} Treating 
both the Sun and aperture as rings rather than disks yields a complete
design equation for multi-disk solar occulters in the SPW approximation:
\begin{equation}
    I_{ap} = I_0 \frac{\Theta_\odot R}{\Theta_{in} h}\mathcal{M}\left(\left[\Theta_{in}-\Theta_\odot\right]\sqrt{\pi\Delta L/\lambda}\right)\left[\mathcal{M}\left(\Delta\Theta\sqrt{\pi\Delta L/\lambda}\right)\right]^{n-2}\mathcal{M}\left(\Delta\Theta\sqrt{\pi(D-L)/\lambda}\right)\,,
    \label{eq:end-to-end-circular-atten}
\end{equation}
which echoes Equation \ref{eq:multiscatter-approximate-I} for the circular case.  Here, $I_{ap}$ is again the 
aperture-averaged intensity of diffracted light; $\Theta_\odot$ is the 
apparent solar radius; $\Theta_in$ is the design acceptance angle of the occulter, i.e. the largest angle from
centerline that the occulter is designed to hide; $R$ is the radius of the aperture; $h$ is the radius of the
final disk of the occulter; $\mathcal{M}$ is defined in Equation \ref{eq:define-mathcal-M}; $\Delta L$ is the 
distance between disks on the occulter; $n$ is the number of disks in the occulter; $L$ is the overall 
length of the occulter; $D$ is the distance 
from the front of the occulter to the aperture; and $\lambda$ is the operating wavelength.  Taking 
$\Delta\Theta \sqrt{\Delta L/\lambda} \ll 1$, and expanding $\mathcal{M}(\gamma)$ to first order,
yields an even simpler design equation, in the sense that most of the transcendental calculations
are eliminated.  
\begin{equation}
    I_{ap} = \frac{I_0\Theta_\odot R}{4^{n-1}\Theta_{in} h }
    \left[1-8\left(\Theta_{in}-\Theta_\odot\right)\sqrt{\Delta L/\pi\lambda}\right]
    \left[1-8\left(\Delta\Theta\sqrt{\Delta L/\pi\lambda}\right)\right]^{n-2}
    \mathcal{M}\left(\Delta \Theta \sqrt{\pi(D-L)/\lambda}\right)
    \label{eq:end-to-end-circular-approx}
\end{equation}
Equation \ref{eq:end-to-end-circular-approx} is a closed-form expression but still contains
the transcendental $\mathcal{C}(\gamma)$ and $\mathcal{S}(\gamma)$ functions, embedded in 
the $\mathcal{M}(\gamma)$ function that forms the final coefficient. 
It is of course possible to approximate even further and eliminate the transcendentals to simplify
the design process.  Ignoring the
advantage of the longer $(D-L)$ throw at the exit of the occulter yields a conservative estimate
with no transcendental functions at all:
\begin{equation}
    I_{ap} \approx \frac{I_0 \Theta_\odot R}{4^{n}\Theta_{in} h}
    \left[1-8\left(\Theta_{in}-\Theta_\odot\right)\sqrt{\Delta L/\pi\lambda}\right]
    \left[1-8\left(\Delta\Theta \sqrt{\Delta L/\pi\lambda}\right)\right]^{n-1}\,.
    \label{eq:no-transcendentals}
\end{equation}
Equation \ref{eq:no-transcendentals} is a complete closed-form solution estimating the 
overall attenuation of an $n$-disk
externally occulted solar coronagraph with prescribed geometry, subject to the limits of the SPW
approximation
as described in Section \ref{sec:SPW-breakdown-wavefront}.  In the condition where 
$\Delta\Theta\sqrt{\Delta L/\pi\lambda}<1$, Equation \ref{eq:no-transcendentals} is always 
conservative compared to the full SPW treatment embodied in
Equation \ref{eq:end-to-end-circular-atten}, and may therefore be used for initial design estimation.

\section{Discussion and Conclusions \label{sec:discussion-conclusions}}

We have developed a simplified theory of occulter performance, suitable for
understanding the behavior of multi-vane linear or corral-baffle occulters (for heliospheric imagers), 
or multi-disk
circular occulters (for externally occulted coronagraphs), in a design regime that
is typical for existing and planned instrumentation.  The successive plane wave (SPW) approximation
represents a ``middle ground'' solution for calculating the diffraction scattering around 
multiple-edge occulters.  Based on the SPW approach, we have identified the ideal envelope
for a multi-edge occulter: a circular-cross section cylinder or ogive, in linear or circular 
geometry respectively.  We have also explained the quasi-achromatic fringes that are 
observed around multi-edge external occulters in modern instrumentation.  By exploring the 
limits of the SPW approach, we have identified a fundamental limit to occulter 
performance
within a prescribed instrument size and angular offset, explaining why high-$n$ solutions 
and smooth occulters do not perform as well as the na\"ive first-order theory or SPW
approximation
predict. Further, we have have developed a formula indicating the highest number of edges that
are useful for a given occulter geometry, before the point
of diminishing returns is reached.  We have also touched on the reasons for the surprisingly
good performance of conical-profile multi-disk coronagraph occulters: even in the 
simple SPW
approximation to the total wave theory, the intermediate disks interact with the 
extended wave field of 
diffracted light near the occulter, despite not protruding into the ray-approximation
beam passing around each disk.

In circular geometry, we have explored the brightness and overall flux of the 
Arago
spot 
formed by coherent interference around an occulter, affirmed the known result that 
the Arago spot is important for single-disk designs, and identified why
the spot is negligible for multi-disk designs.  In particular, in multi-disk designs
the Arago spot is attenuated by the action of the $n-1$ disks preceding the last one, 
and is therefore reduced by a large factor in brightness compared to a single-disk system. 
We have also discussed the incoherent concentration
of diffracted light by the circular geometry of the occulter, even outside the Arago spot.  
We have identified analytically the brightness pattern expected across the aperture of 
a multi-disk externally occulted coronagraph.  Finally, we have generated
conservative design estimation 
formulae that permit iterative design of novel instruments via closed-form analytic
expressions
to estimate occulter performance, along with slightly more complex formulae that may be 
applied to model the stray light from a chosen design.  \new{The conservatism of the 
estimation formulae is important, because of the known sensitivity of multi-disk occulter
performance to small misalignments.  
These formulae do not require explicit simulation of the electromagnetic field (or, 
equivalently, numeric integration of the
Kirchoff integral), yet still
capture enough of the physics of occulter behavior to both advance 
intuition and 
yield predictive power for new designs.}

Occculter design is only one element of either heliospheric imager or coronagraph
design.  Even fully externally occulted designs require considering scattering 
in the remainder of the instrument including the focusing optics and detector 
system, which scatter the ``bright line'' or ``bright ring'' from the external
occulter into broader stray light patterns at the focal plane.  
Further, both 
types of instrument view a faint object -- the Sun's K 
corona -- against
a much brighter background comprising a mix of sky brightness, the solar F corona, and 
the starfield \new{in} addition to any instrumental stray light.  Coronagraph images 
must thus be 
post-processed to remove all of these sources of brightness, in order to reveal the 
K corona; precise photometry and careful post-processing are therefore required regardless
of the level of instrumental stray light.  

Even a highly simplified wave theory such as the successive plane wave treatment has 
predictive and explanatory power for the behavior of multi-disk occulters.  By exploring 
the limits of this approximation, we have been able to explain and quantify
several aspects of 
occulter behavior -- achromatic fringing, the limits of multi-disk performance, and the 
\textit{lack} of degradation, in circular geometry, from the Arago spot in multi-disk coronagraphs.
Further, the
successive-plane-wave treatment has sufficiently
well defined limits of applicability, that it may be used
to inform future instrument designs in both linear and circular geometry.

\begin{acknowledgments}
This work was partially funded by the NASA/NOAA Space Weather Next 
program, Contract No. 80GSFC23CA056x, and partially funded internally by Southwest
Research Institute under its Internal Research \& Development program.  
The authors gratefully acknowledge helpful comments from several members
of the coronagraph instrument community, including C. Aime, R. Howard, and A. Thernisien.

\end{acknowledgments}

\bibliographystyle{aasjournal}
\bibliography{occulter-design}{}

\end{document}